\documentclass[structabstract]{aa}

\usepackage{amsmath}
\usepackage{natbib}
\usepackage{epsfig}
\usepackage{txfonts}
\usepackage{multirow}
\usepackage{hyperref}

\newcommand{\cmms}{\ensuremath{\mathrm{cm}^{-2}~\mathrm{s}^{-1}}}
\newcommand{\gcc}{\ensuremath{\mathrm{g~cm^{-3}}}}

\newcommand{\nuc}[2]{\ensuremath{\mathrm{^{#1}#2}}}
\newcommand{\ye}{Y_{\rm e}} 
\newcommand{\nue}{\nu_{\rm e}} 
\newcommand{\nuebar}{\bar \nu_{\rm e}} 
\newcommand{\numu}{\nu_{\rm \mu}}

\bibpunct{(}{)}{;}{a}{}{,}

\begin{document}
\bibstyle{default}

\title{Probing thermonuclear supernova explosions with neutrinos}

\titlerunning{SN Ia neutrino emission}

\author{
        A. Odrzywolek\inst{1}\and
        T. Plewa\inst{2}
       }

\authorrunning{A. Odrzywolek \& T. Plewa}

\offprints{A.~Odrzywolek \email{andrzej.odrzywolek@uj.edu.pl}} 
       
\institute{
  Marian Smoluchowski Institute of Physics,
  Jagiellonian University,
  Reymonta 4, 
  30-059 Cracow, Poland \and
  Department of Scientific Computing,
  Florida State University,
  Tallahassee, FL 32306, U.S.A.
}

\date{Received / Accepted}      

\begin{abstract} 
{}
{We present neutrino light curves and energy spectra for two
representative type Ia supernova explosion models: a pure
deflagration and a delayed detonation.}
{We calculate the neutrino flux from $\beta$ processes using nuclear statistical equilibrium
abundances convoluted with approximate neutrino spectra of the
individual nuclei and the thermal neutrino spectrum (pair+plasma).}
{Although the two considered thermonuclear supernova explosion
scenarios are expected to produce almost identical electromagnetic
output, their neutrino signatures appear vastly different, which allow an
unambiguous identification of the explosion mechanism: a pure
deflagration produces a single peak in the neutrino light curve, while the
addition of the second maximum characterizes a delayed-detonation.
We identified the following main contributors to the neutrino signal:
(1) weak electron neutrino emission from electron captures (in
particular on the protons \nuc{55}{Co} and \nuc{56}{Ni}) and numerous
$\beta$-active nuclei produced by the thermonuclear flame and/or
detonation front, (2) electron antineutrinos from positron captures on
neutrons, and (3) the thermal emission from pair annihilation. We
estimate that a pure deflagration supernova explosion at a distance of
1~kpc would trigger about 14 events in the future 50 kt liquid
scintillator detector and some 19 events in a 0.5 Mt water
Cherenkov-type detector.}
{While in contrast to core-collapse supernovae neutrinos carry only a
very small fraction of the energy produced in the thermonuclear
supernova explosion, the SN Ia neutrino signal provides information
that allows us to unambiguously distinguish between different possible
explosion scenarios. These studies will become feasible with the next
generation of proposed neutrino observatories.}
\end{abstract}


\maketitle

\section{Introduction\label{s:introduction}}

The origins of type Ia supernovae (SN Ia) remain one of the major
unsolved problems of stellar evolution
\citep{hoeflich+02,kuhlen+06,piro08,zingale+09}. The commonly accepted
theoretical framework considers an explosion scenario in which a
massive white dwarf slowly gains mass in the process of accretion from
a non-degenerate companion
\citep{whelan+73,yoon+03,han+04,meng+10}. Alternatively, the
degenerate matter might be ignited in the process of a violent merger
of binary white dwarfs \citep{iben+84,webbink84,han98}. The latter
channel might be a dominant source of thermonuclear events in early
type galaxies \citep{gilfanov+10,wang+10b}, while there is no
consensus as to which evolutionary process dominates in other
environments \citep{scannapieco+05,raskin+09,ruiter+09,schawinski09}.

Our progress toward understanding these events is hampered by the
relatively low luminosity of their progenitors, and to date the
evidence is largely circumstantial and exclusively indirect
\citep{ruiz-lapuente+04,badenes+07,schawinski09,gilfanov+10}. This
stays in contrast with numerous identifications of core-collapse
progenitors \citep[][and references therein]{smartt09,leonard09}.
Furthermore, the nature of the explosion process is very uncertain,
though it is commonly accepted that the energy source of the explosion
is a thermonuclear burn \citep{hoyle+60}. For a single-degenerate channel, the
nuclear fuel is expected to burn first subsonically \citep{nomoto+76}
with a likely transition to detonation at a later time
\citep{Khokhlov_tau_NSE_2,woosley+94}. It is much less clear what the
ultimate fate of the merger is
\citep{hachisu+86,saio+85,yoon+07,pakmor+10}, and perhaps additional
routes to an explosion are admissible \cite[][and references
therein]{podsiadlowski+08,podsiadlowski10}. These questions along with
the role that SN Ia play in studies of the early universe
\citep{sandage+93,riess+98,phillips05,wood-vasey+07,ellis+08,riess+09,kessler+09}
motivate our search for additional sources of information about
thermonuclear supernovae, and in particular about the explosion
process.

Neutrinos are a proven source of  information about
astrophysical objects and phenomena, such as the Earth
\citep{TriesteWorkshop,2005Natur.436..499A,Geoneutrino2005}, and
engineering systems such as nuclear power plants
\citep{1742-6596-136-2-022008,Lhuillier2009112,Learned2005152,Geo_nkorea}.
The Sun is one of the best-studied astrophysical neutrino sources
thanks to its proximity and constancy of the $\nu_e$ flux
\citep{bahcall89}. Solar neutrino studies were first conducted using
radiochemical detectors \citep{HOMESTAKE1,GALLEX1} and more recently
also in real-time
\citep{2008PhLB..658..101B,2008PhRvL.101i1302A,SK_solar8,SNO_solar1}. For
contemporary non-solar neutrino experiments, the solar neutrino signal
caused by the dominant reactions (pp, \nuc{8}{B}) constitutes somewhat
undesirable background. However, supernova SN 1987A
\citep{1989ARA&A..27..629A} has been clearly observed in neutrinos in
many detectors \citep{IMB_sn1,SK_sn3,LSD_sn1,Baksan_sn3} despite its
nearly extragalactic distance ($\sim$50~kpc). The event has been the
main trigger for intensive theoretical studies and modeling in the
recent years \citep{immler+07,nakahata+07} while a possibility of
neutrino detection and obtaining neutrino energy spectra from
core-collapse supernovae \citep{SN_neutrinos,0004-637X-590-2-971}
attracted constant attention of theorists
\citep{2008arXiv0810.1959K,Mirizzi,2005PhRvL..95q1101A,2005JCAP...04..002F}
and stimulated experimental developments
\citep{2001hep.ex...10005S,GigatonArray}. Neutrino detection is a
mature field of research nowadays. For instance, a stellar core-collapse
at a distance $< 4$ kpc will produce a signal strong enough to saturate
the Super-Kamiokande detector \citep{Nakahata_SN1987A-20th}. Therefore,
it is natural to consider the detectability of neutrinos from previously
ignored sources, including thermonuclear supernova events.

As originally suggested by \cite{DarkSupernovae}, the neutrino signal
produced by the thermonuclear deflagrations offers direct insight into
the explosion process. Clearly, such observations would be extremely
helpful in directing future SN Ia research and may possibly allow for
distinguishing between various stellar evolution and explosion
scenarios. A striking differences between neutrino emission from
deflagrations and delayed detonations has been noted by
\cite{DarkSupernovae}. More recently, in a series of articles Kunugise
\& Iwamoto \citep{iwamoto+06,kunugise+07} studied the $\nu_e$ light
curve and spectra from the standard W7 explosion model
\citep{Nomoto_W7} and discussed the detectability of this type of event by the
Super-Kamiokande detector. We aim to extend those early studies to
recent multi-dimensional thermonuclear supernova explosion models. We
obtain supernova neutrino light curves and energy spectra for pure
deflagration and delayed detonation explosion models. We show that the
predicted neutrino signatures are markedly different in those two
cases and can be used to identify the explosion mechanism.

\section{Neutrino emission from thermonuclear supernovae\label{s:emission}}

Neutrino emission from a type Ia supernova is considered negligible in
most of the thermonuclear explosion models because the weak interaction
rates are too slow compared to the hydrodynamic timescale
\cite[see][Sect. 9.1]{arnett96} and the matter is essentially
completely transparent to neutrinos. However, it is conceivable that
if the amount of the energy emitted via neutrinos is significant
compared to the energy produced in the thermonuclear burning, the
neutrino cooling may play an important role in the explosion dynamics.
In either case, neutrinos may provide important insights into the SN
Ia explosion mechanism.

Neutrino emission from the existing SN Ia explosion models can be
computed by post-processing snapshots of the hydrodynamical
simulations. For the thermal neutrino emission this is a
straightforward procedure because the neutrino spectrum only depends on the
temperature and the (electron) density of the plasma. For weak nuclear
processes, we have to know the isotopic composition of the
plasma. Given the current computational resources, it is not feasible to
include large nuclear reaction networks in multidimensional explosion
model. The situation, however, is not completely hopeless because the
hottest regions associated with thermonuclear flames and detonations,
which is also where the neutrino emission is expected to be relatively high, are
in the nuclear statistical equilibrium (NSE)
\cite[see][Sect. 7.2]{clayton84}. Under NSE conditions, isotopic
abundances are determined solely by the thermodynamic properties of
the plasma. Therefore, in the most important regions of the exploding
star, we are again able to post-process models and compute required
abundances. Once the isotopic composition is known, computing a neutrino
emission is relatively straightforward \citep{kunugise+07}.

In NSE, the isotopic composition of the matter is fully determined by
the density, temperature, and electron density of the plasma
\citep{CliffordTayler_2, CliffordTayler}. The NSE conditions are
characterized by
\begin{enumerate}
\item a very high temperature to break-up the most strongly bound
nuclei;
\item an evolutionary timescale long enough to
allow for re-arranging of nucleons into equilibrium nuclei
{em via} strong/electromagnetic interactions.
\end{enumerate}
These conditions can be found in the iron cores of pre-supernova stars,
during core-collapse, and last but not least, during thermonuclear
burn in type Ia supernovae. More recently, protoneutron star evolution
and accretion-induced collapse recently has been analyzed from this
point of view by \cite{Arcones2010}.

For completeness we will discuss shortly
the major properties of the considered neutrino emission
processes. Model neutrino spectra are computed with help of the PSNS
code \citep{PSNS}.

\subsection{Sources of neutrinos}

\subsubsection{Thermal processes}

Three ''classic'' neutrino processes,
\begin{subequations} 
\begin{equation}
\label{pair} e^{-} + e^{+} \to \nu_{e, \mu, \tau} + \bar{\nu}_{e, \mu, \tau}
\end{equation}
\begin{equation}
\label{plasma} \gamma^\ast_{L,T} \to \nu_{e, \mu, \tau} + \bar{\nu}_{e, \mu,
\tau} \end{equation} \begin{equation} \label{photo} \gamma + e^{-} \to e^{-} +
\nu_{e, \mu, \tau} + \bar{\nu}_{e, \mu, \tau} \end{equation}
\end{subequations}
are the major source of the so-called thermal neutrinos \citep{Itoh_I,
Schinder, Esposito}: annihilation of the $e^{+}e^{-}$ pairs into
neutrinos (Eq.~\eqref{pair}, \citealt{OMK3}); plasmon decay,
(Eq.~\eqref{plasma}, \citealt{BraatenPRL, BraatenSegel}), and
photoemission (Eq.~\eqref{photo}, \citealt{Dutta}). Emissivity and
spectra of these neutrinos are uniquely determined by the plasma
temperature and electron density. All flavors of the neutrinos are
produced in these processes: $\nu_e, \bar{\nu}_e, \nu_{\mu},
\bar{\nu}_{\mu}, \nu_{\tau}, \bar{\nu}_{\tau}$. Following the standard
theory of electroweak interactions, the fluxes for all flavors are quite
similar, yet some differences exist between the electron and
$\mu/\tau$ flavors. Additionally, because of the parity violation,
neutrino and antineutrino energies are not equal under the degenerate
conditions considered here \citep{Odrzywolek_plasmaneutrino,OMK3}.

Pair annihilation neutrino fluxes and spectra were calculated
according to \cite{OMK3}. This approach is superior to both the
\cite{Itoh_VII} method, which is typically used in stellar evolution
calculations (because the neutrino flavors are not summed up) and the
\cite{Bruenn1985, BurrowsThompsonPairs} method that is used for core-collapse
supernova modeling (because the electron rest mass is neglected).

The plasma neutrino flux and spectrum were calculated according to
\cite{Odrzywolek_plasmaneutrino}. Procedures were tested against the
\cite{Itoh_IV}, \cite{Itoh_VI}, and \cite{Itoh_VIII} tables
(calculated using slightly different dispersion relations for
plasmons) with reasonable agreement, and also against the recent
calculations of \cite{KantorGusakov}. In the latter case, the results
agree up to the machine precision.

The photoneutrino process and thermal processes of a lesser
importance (e.g.\ neutrino bremsstrahlung, cf.\ \citealt{Hansel}) were
omitted in our calculations, because of the lack of relevant results on the
neutrino spectrum. This may lead to a negligible underestimate of the
thermal neutrino flux.

\subsubsection{Weak nuclear processes}
      
Weak processes, namely electron/positron captures on both nucleons and
nuclei and $\beta^\pm$ decays are extremely important in the
astrophysical environments. They are essential ingredients of,
e.g.,\ massive star evolution (especially pre-supernova phase,
\citealt{2009AcPPB..40.3063K}), core-collapse supernovae, and
thermonuclear explosions: x-ray flashes, novae and SN Ia. Weak nuclear
neutrino processes usually work in the cycles such as:
\begin{subequations} 
\label{weak}
\begin{equation} \begin{aligned} e^{-} + (A&,Z)
\longrightarrow (A,Z\!-\!1) + \nu_e\\ &\uparrow \qquad \qquad \downarrow \\
\bar{\nu}_e + e^{-} + (A&,Z) \longleftarrow (A,Z\!-\!1) \end{aligned}
\end{equation} \begin{equation} \begin{aligned} e^{+} + (A&,Z\!-\!1)
\longrightarrow (A,Z) + \bar{\nu}_e\\ &\uparrow \qquad \qquad \quad \downarrow
\\ \nu_e + e^{+} + (A&,Z\!-\!1) \longleftarrow (A,Z) \quad, \end{aligned} \end{equation}
\end{subequations}
and the total number of emitted neutrinos per nucleus is usually not
equal to 1, in contrast to terrestrial beta decays and electron
captures.

One of the most important motivations for including the weak nuclear
rates was a search for nuclei producing $\nuebar$ or $\nue$, which leads to a very strong signal
in the detectors (in analogy to Solar \nuc{8}B neutrinos). 
These nuclei must meet three conditions: (1) they have to be
abundant in NSE, (2) they need posses very high $\beta$ or a very high capture
rate, and (3) they need to emit energetic $\nue$ or $\nuebar$ with energies above, say,
10-15 MeV. Unfortunately, an inspection of the Figs.~\ref{nuHR-n7-ES4},\ref{nuebarHR-y12-IBD2} 
and Table~\ref{t:nu_events} reveals no such nuclei in our
study. A strong degeneracy during the initial stage of the deflagration
enhances transitions with relatively high-energy neutrinos (we thank
G. Fuller for pointing out this important aspect to us). For some nuclides,
e.g.\ \nuc{57}{Zn}, \nuc{54}{Cr}, and \nuc{28}{P}, the
average neutrino energy $\langle \mathcal{E}_{\nu_e} \rangle$ reaches
15~MeV. The NSE abundance and therefore the neutrino flux from
these nuclides is negligible (cf.\ Fig.~\ref{nuHR-n7-ES4}). The
nucleus producing the highest elastic scattering event rate is
$^{55}$Co, but equally important are electron captures on protons. The
case of $^{54}$Co, with a quite high average neutrino energy
($\approx$9~MeV) is very interesting and deserves a more detailed
analysis.

Some of the nuclei also produce  relatively energetic antineutrinos,
e.g.\ $\langle \mathcal{E}_{\nu_e} \rangle \approx 6 $ MeV for $^{56}$V
and $^{58}$V during the deflagration and detonation stages. The
corresponding flux, however, is low compared to thermal (pair) and
$e^{+}(n,p)\bar{\nu}_e$ electron antineutrinos fluxes. We conclude
that the $\beta$ processes involving nuclei provide only a negligible
contribution to the $\bar{\nu}_e$ flux.

While the energy loss rate as well as the decrease of the electron
fraction because of weak processes were extensively studied in the past
\cite[][and references therein]{FFN_I, FFN_II, FFN_III,
1994ADNDT..56..231O, 1994ApJ...424..257A, 1994ApJS...91..389A,
1999NuPhA.653..439C, 2000NuPhA.673..481L, 1999ADNDT..71..149N,
2009ADNDT..95...96S, 2010NuPhA.848..454J}, relatively little is known
about the combined energy spectrum of these neutrinos
\citep{PhysRevC.64.055801,2009PhRvC..80d5801O}. Typically, the
spectrum is integrated in advance and the results are tabulated. This
approach saves both computer memory and computing time. To restore
information about the spectrum, a simple parameterization (e.g.\ the
Fermi-Dirac distribution) is assumed \cite[see, e.g.][]{Pons}. We
employ a similar method here. However, some fine details of the
nuclear structure reflected in the neutrino spectrum are lost when
using this approach. In certain conditions, this may lead to a serious
underestimate of the neutrino signal, especially in the high-energy
($\mathcal{E}_{\nu} > 10$~MeV) tail. With this in mind, our results
provide a lower detection threshold for the neutrino
signal. Furthermore, some newest results suggest an upward revision of
the crucial \nuc{55}{Co} electron capture rate by up to two orders of
magnitude \citep{Nabi_Co55_2008}. These findings apparently are
in conflict with the nucleosynthesis results though, in particular with the
observed degree of neutronization of the ejecta
\citep{1997NuPhA.621..467N, 1993nuco.conf..569I, Thielemann198467,
1999ApJS..125..439I}.

Our calculations of the weak nuclear neutrino emission proceed as
follows.  In contrast to the thermal neutrino emission, the
contribution from weak nuclear processes to the neutrino signal
cannot be calculated solely based on the thermodynamic properties of
matter. These calculations in general require detailed knowledge of the
isotopic composition. Typically, the composition is a result of the
long and complicated history of the astrophysical object. Because the
electron fraction has not been calculated consistently in the adopted
explosion models, we assume $\ye = 0.5$. This value corresponds to the
initial electron fraction of the progenitor with 50/50 carbon/oxygen
composition mix used in the explosion calculations.\footnote{In more
realistic progenitor models, $\ye$ should be slightly below 0.5 because of
core burning before the explosion \citep{2008ApJ...673.1009P} and/or
variation in the initial chemical composition of the progenitor star
on the main-sequence \citep{timmes+03}.}. In more realistic models,
the electron neutrino emission would result in decreasing $\ye$. For
example, the NSE abundance of the \nuc{55}{Co} nucleus, which
significantly contributes to the $\nue$ flux, decreases rapidly for
$\ye<0.5$.

The remaining required information about the matter density, $\rho$,
and the temperature, $T$, is obtained from the actual explosion model. We
consider only regions where the NSE state can be established on a
timescale shorter than the explosion timescale.  The NSE timescale can
be approximated as \citep{Khokhlov_tau_NSE_1, Khokhlov_tau_NSE_2}
      \begin{equation}
      \label{time-to-nse}
      \tau_{\mathrm{NSE}} \sim \rho^{0.2} e^{179.7/T_9 - 40.5} \; \mathrm{s}.
      \end{equation}
For the reference NSE threshold temperature, $T_{\mathrm{NSE}}=5
\times 10^9$ K ($T_9=5$, kT$\approx$0.432~MeV) , adopted after
\cite{kunugise+07} and the characteristic density of $\rho = 10^9
\gcc$, the NSE timescale is, $\tau_{\mathrm{NSE}}\approx 0.66$ s, and
is shorter than than the explosion timescale,
$\tau_{\mathrm{exp}}\approx 1$ s.

To estimate the sensitivity of the results to the assumed NSE threshold
temperature, we performed several additional calculations with the
threshold temperature $T_9 = 6$ (kT$\approx$0.517~MeV,
$\tau_{\mathrm{NSE}}\sim 10^{-3}$~s). This resulted in a reduction of the total
neutrino flux by a few percent. The remaining non-NSE zones
were omitted from the weak neutrino emission calculations\footnote{Those
regions produce neutrinos from decaying beta-unstable nuclides,
e.g. \nuc{56}{Ni}. This process does not depend on temperature.}. Their
contribution remains unknown at present, but it is unlikely to be
important.

For zones with $T>T_{\mathrm{NSE}}$, the NSE abundances were
calculated using an 800 isotope network up to \nuc{97}{Br}
\citep{2009PhRvC..80d5801O}. From the NSE abundances, we selected
nuclei (188 nuclides) for which weak rates have been tabulated by
\cite{FFN_I, FFN_II, FFN_III}. Model energy spectra for neutrinos from
electron captures on protons and for antineutrinos from positron
captures on neutrons and neutron decay were calculated using
\begin{eqnarray}
\label{approx_spectrum}
\frac{d R_{\nu}}{d \mathcal{E}_{
\nu}} & = & \left(\frac{\ln{2}}{m_e^5}\right) r_{eff} \Theta\left(\pm \mathcal{E}_\nu \mp Q_{eff}-m_e\right) \\
 & & \frac{\mathcal{E}_\nu^2 (\pm \mathcal{E}_\nu \mp Q_{eff}) \sqrt{ (\mathcal{E}_\nu-Q_{eff})^2-m_e^2} }{ 1 + e^{(\mathcal{E}_\nu-Q_{eff} \mp \mu)/kT} },
\nonumber
\end{eqnarray}
where $R_{\nu}$ is the particle production rate per unit volume and
time, $\mathcal{E}_\nu$ is the neutrino energy, $r_{eff}$ and
$Q_{eff}$ describe adopted parameterization \cite[see][for
details]{PhysRevC.64.055801}, $\Theta$ is the unit step function,
upper and lower sign correspond to captures and decays, respectively,
and the other symbols have their usual meanings. To account for
positron captures ($\epsilon^+$) and $\beta^+$ decays, one simply needs
to change the sign of $\mu$ (the electron chemical potential including
rest mass) in Eq.~\eqref{approx_spectrum}. The neutrino spectra were
calculated using Eq.~\eqref{approx_spectrum} with the effective
Q-values and effective rates \citep{PhysRevC.64.055801,kunugise+07}
with additional switching between capture and decay
\citep{2009PhRvC..80d5801O}. The above procedure reproduces neutrino
fluxes and average neutrino energies of the original tabulated values
at the FFN grid points. Between grid points, we used a bilinear
interpolation of the effective rates and Q-values \citep{FFN_IV}.
The electron chemical potential required in Eq.~\eqref{approx_spectrum}
was computed separately with a precision better than $1\times
10^{-12}$.

\subsection{Representative SN Ia explosion models}

For the neutrino explosion diagnostic analysis, we selected two
representative explosion models from our database \citep{Plewa_1}: a
pure deflagration, n7d1r10t15c, and a delayed detonation, Y12. Both
models were obtained for a standard carbon/oxygen Chandrasekhar mass
white dwarf. A slightly modified flame capturing method of
\cite{khokhlov95} was used to follow a deflagration, and we used a 13-isotope 
alpha-network to directly compute the energetics of the detonation
wave. Both models are relatively energetic with explosion energies
between $\approx 0.97$~B (1~Bethe $\equiv$ 1~B = $1\times 10^{51}$ ergs) for the
pure deflagration and $\approx 1.36$~B for the
delayed-detonation.

\subsection{Detailed analysis of the neutrino emission}

For the selected explosion models, we computed the neutrino emission
resulting from pair annihilation, plasmon decay and weak nuclear
processes. The results are presented in the form of emissivity maps
and total fluxes. Additionally, we provide time-dependent neutrino
energy spectra in numerical form (see online materials). Following
practice known from core-collapse supernova studies, we show
individual neutrino emission light curves for electron neutrinos
($\nu_e$), electron antineutrinos ($\bar{\nu}_e$) and the average of
the remaining four muon and tau neutrinos ($\nu_\mu$). The latter are
produced exclusively in thermal processes, as long as we neglect
neutrino oscillations. The electron neutrino ($\nue$) flux is dominated
either by electron captures on protons and iron group nuclei
\footnote{Especially \nuc{55}{Co} and \nuc{56}{Ni}.} (when the
burning is the most intense) or by pair annihilation (otherwise).

Electron antineutrinos ($\nuebar$) are produced mainly in the pair
process and through positron captures on neutrons. Heavy nuclei
($\beta^-$ decays and $e^{+}$ captures) do not significantly contribute to the total
$\nuebar$ flux. Muon and tau neutrinos are produced in
much smaller quantities only in the thermal processes, and one may
expect that actually more $\mu/\tau$ neutrinos are produced owing to flavor
conversion between source and detector (see Fig.~3 in
\cite{kunugise+07}). Plasmon decay is almost negligible because of the low
densities, and the low energy of the emitted neutrinos ($\sim$few keV,
\citealt{Odrzywolek_plasmaneutrino}) makes their detection essentially
impossible.

\subsubsection{Pure deflagration model}
%
%
%
\begin{figure*}[t!]
  \centering
  \includegraphics[width=0.95\textwidth]{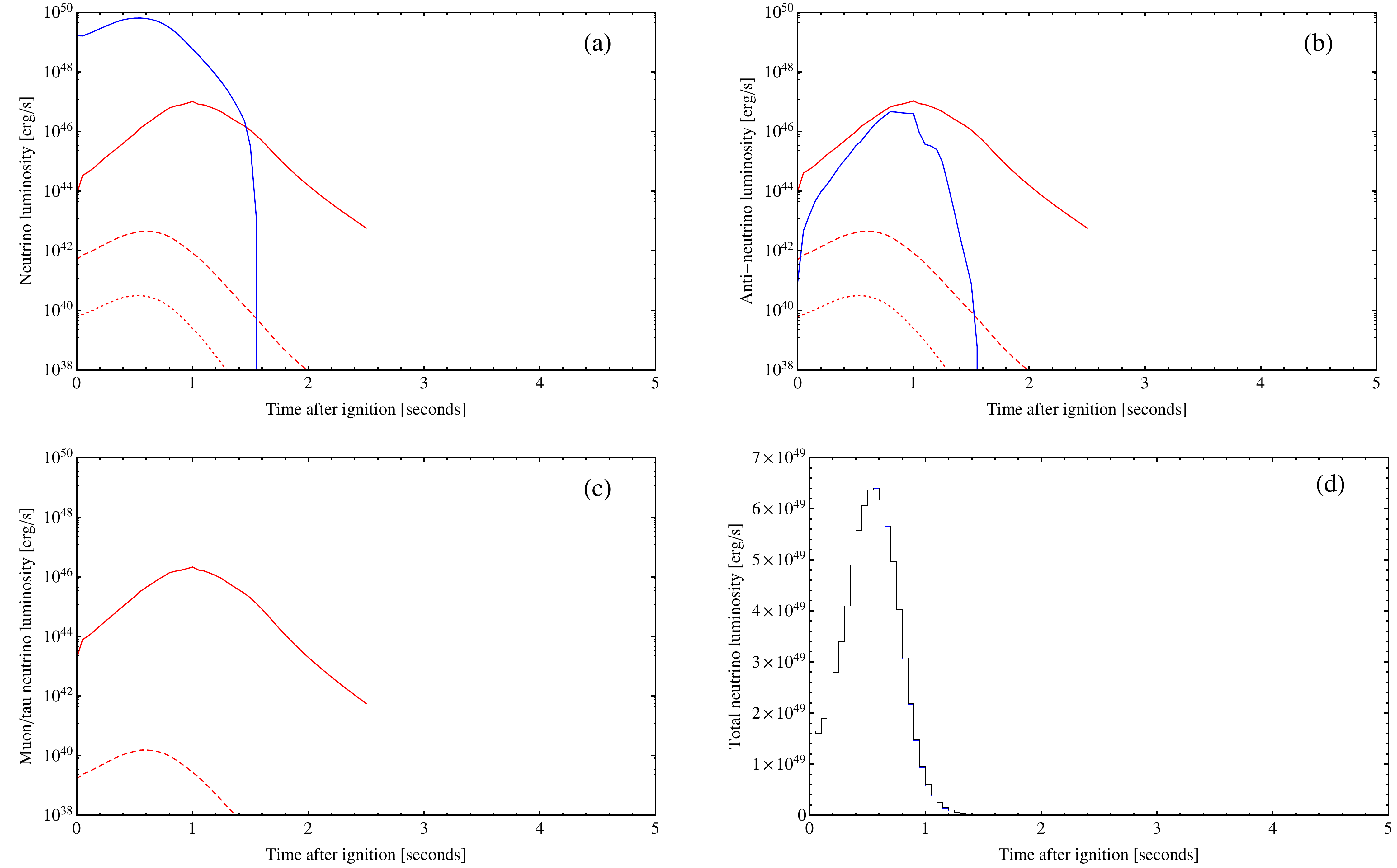}
  \caption{Model neutrino luminosities of the pure deflagration n7d1r10t15c.
    (a) electron neutrinos, $L_{\nu_e}$;
    (b) electron anti-neutrinos, $L_{\bar{\nu}_e}$;
    (c) $\mu$ and $\tau$ neutrinos, $L_{\bar{\nu}_x}$;
    (d) total flux.
    In each panel we show the contribution of weak (solid blue,
    Eq.~\eqref{weak}), pair annihilation (solid red, Eq.~\eqref{pair}), transverse
    (dashed red) and longitudinal (red dotted)
    plasmon decay (Eq.~\eqref{plasma}).}
  \label{Lnu_n7}
\end{figure*}
\begin{figure}
  \centering
  \includegraphics[width=0.5\textwidth]{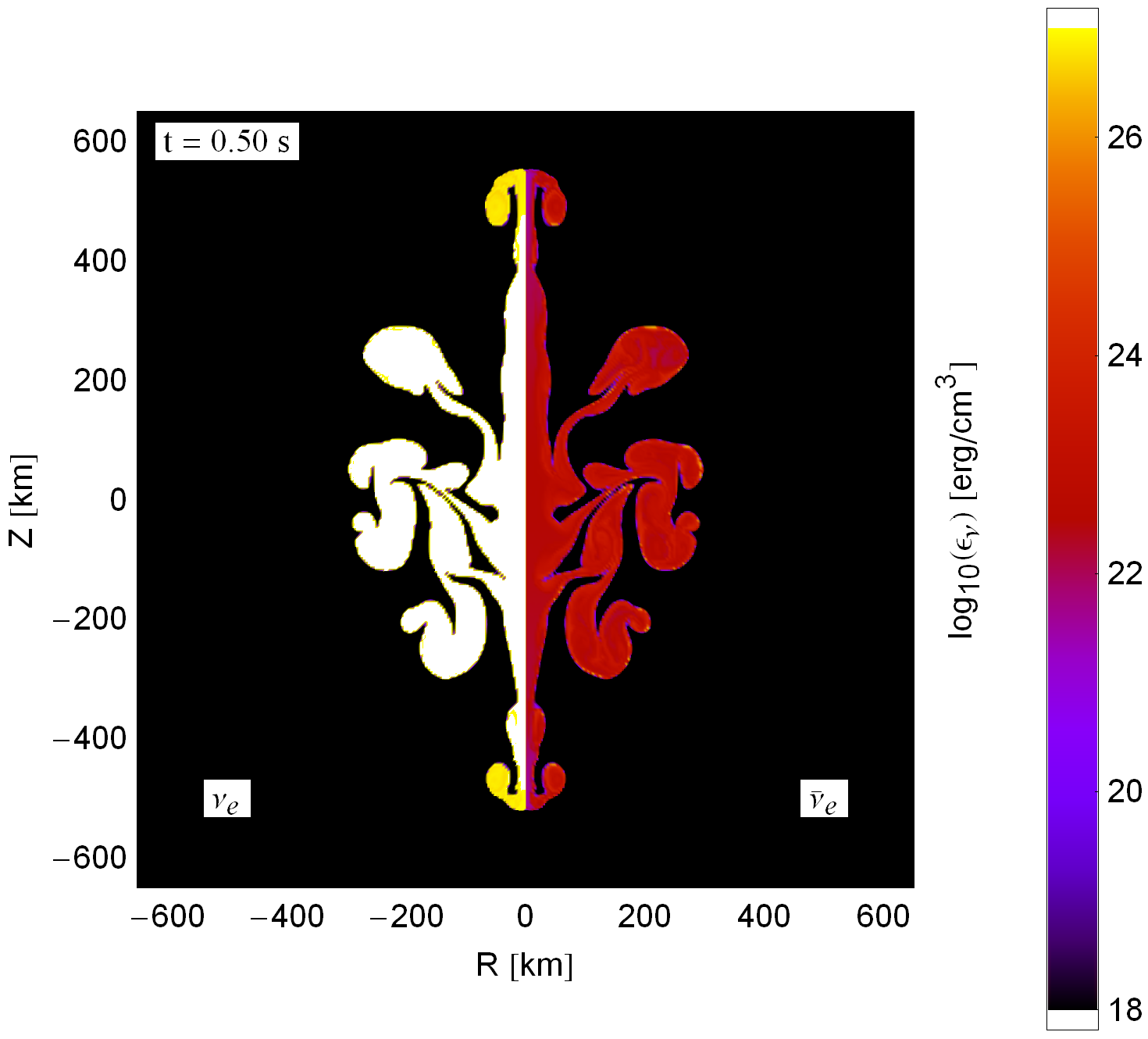}%
  \caption{Maps of the neutrino emissivity in the pure deflagration
 model at $t=0.5$~s (i.e.\ near the peak of the neutrino emission).
 (left segment, $R < 0$ km) $\nu_e$; (right segment, $R > 0$ km)
 $\bar{\nu}_e$.}\label{Lnu_n7_map_deflagration}
\end{figure}

Pure deflagrations produce neutrino emission with a single maximum
(because an explosion involves only one stage), and nuclear burning takes
$\approx 1$ second. We calculate the total neutrino flux
(Fig.~\ref{Lnu_n7}d) as the sum of thermal and weak components. The
evolution is slower compared to a detonation (see below), and in this case
therefore neutrino cooling processes are given more time compared to a
detonation. Moreover, a larger volume is involved in neutrino cooling in
deflagration compared to a ``failed'' case, Y12 (cf.\
Fig.~\ref{Lnu_n7_map_deflagration} versus
Fig.~\ref{Lnu_y12_map_deflagration}). Overall the neutrino luminosity is
much higher compared to Y12 model and reaches $1.92 \times 10^{50}$
erg/s, almost one order of magnitude larger compared to the first-peak
luminosity of the Y12 model ($1.1 \times 10^{49}$ erg/s). The total energy
radiated in neutrinos is 0.04~B, five times
more than for Y12 ($0.008$~B), but still small compared to the overall
explosion energy of $\approx 1$~B.

The temporal evolution of the neutrino emission in the deflagration model
is shown in Fig.~\ref{Lnu_n7}a ($\nu_e$), Fig.~\ref{Lnu_n7}b
($\bar{\nu}_e$), Fig.~\ref{Lnu_n7}c ($\nu_\mu$), with the total
neutrino luminosity shown in Fig.~\ref{Lnu_n7}d. Overall, the emission
varies smoothly in time and we notice only very small emission
fluctuations. Even though the flame is geometrically very convoluted
(Fig.~\ref{Lnu_n7_map_deflagration}), the neutrino emission is
produced in regions of nearly identical density and temperature. We
found that most (99\%) of the NSE neutrino flux is produced for
$T_{NSE}<T_9<10$ and $8.9<\log_{10} \rho<9.3$.  At the peak neutrino
emission, only 3\% of the total white dwarf mass is emitting
neutrinos.

We note that the model neutrino emission obtained in our axisymmetric
deflagration is very similar to that of the spherically symmetric
model W7 \citep{Nomoto_W7,DarkSupernovae, kunugise+07}. This suggests
that the neutrino emission from pure deflagrations may have a generic
form. To verify this impression, we computed the neutrino light
curves for two other deflagration models presented by \citep{Plewa_1},
n11d2r10t15a and n11d2r20t20b. In both cases the neutrino emission
displayed very similar characteristics to W7 and the deflagration
model analyzed in detail here. The generic form of the emission also
implies that \emph{neutrinos may provide no information helpful for
separating between various scenarios of pure deflagrations}.
%
%
%
\subsubsection{Delayed-detonation model}
%
%
%
In contrast to the pure deflagrations, the delayed-detonation class of models
produces multi-peak neutrino emission.
\begin{figure*}[t!]
  \centering
  \includegraphics[width=0.95\textwidth]{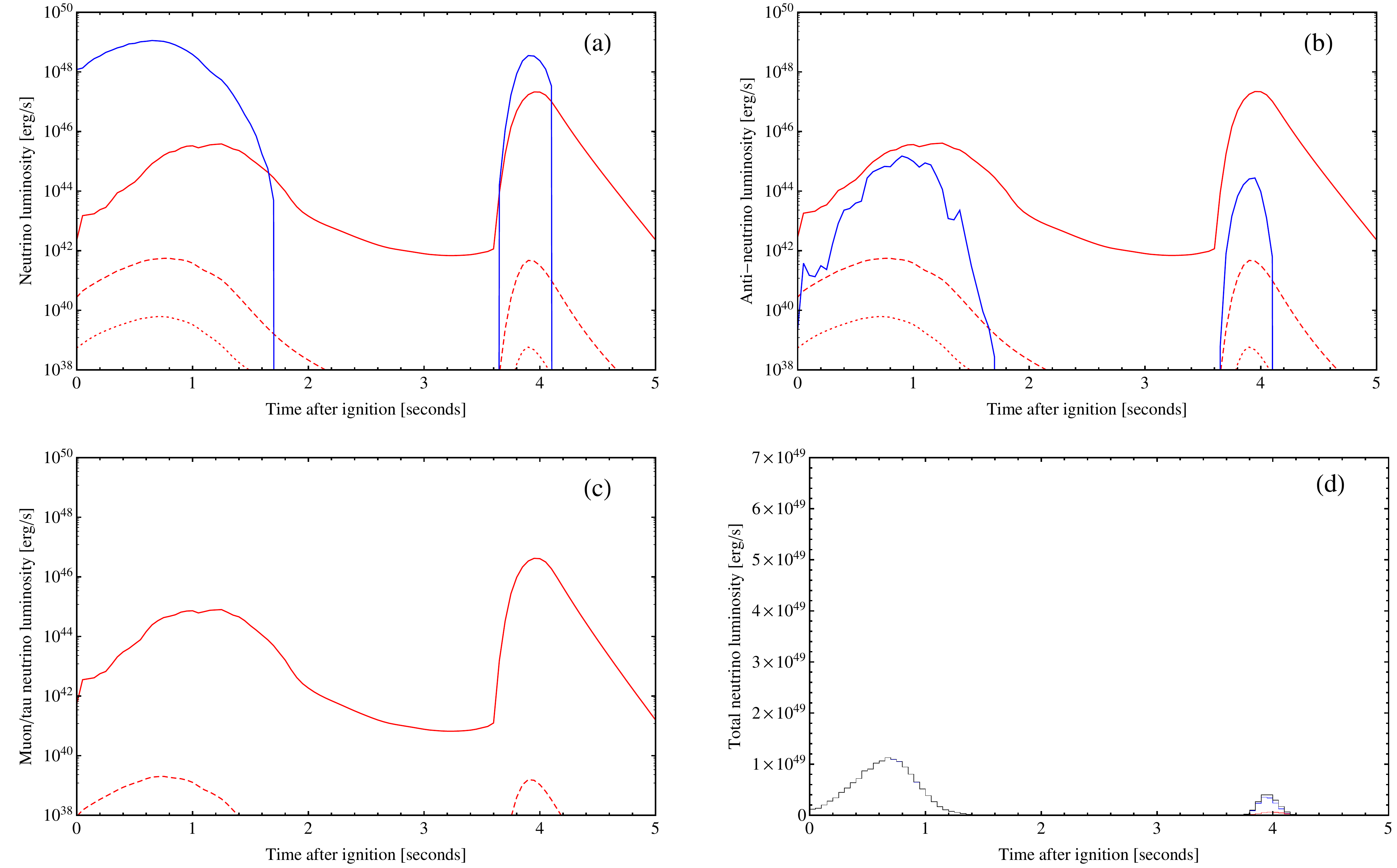}
  \caption{Model neutrino luminosities of the delayed detonation Y12.
    (a) electron neutrinos, $L_{\nu_e}$;
    (b) electron antineutrinos, $L_{\bar{\nu}_e}$;
    (c) $\mu$ and $\tau$ neutrinos, $L_{\bar{\nu}_x}$;
    (d) total flux.
    The color and line-style coding is identical to that in Fig.~\ref{Lnu_n7}.}
  \label{Lnu_y12}
\end{figure*}
The two
distinct neutrino emission maxima caused by the initial deflagration stage
and delayed detonation can be clearly discerned (Fig.~\ref{Lnu_y12}). 
The deflagration peak is
completely dominated by the $\nu_e$ emission from the electron captures.
The detonation peak, while still dominated by the weak nuclear processes,
includes a significant fraction of the thermal emission. Actually, pair
annihilation dominates after end of rapid detonation stage and form an
exponentially decaying tail. This is the result of the efficient neutrino
cooling in the large volume of the former white dwarf that is overrun by the
detonation wave (cf.\ Fig.~\ref{Lnu_y12_map_detonation}).

\begin{figure}[hb!]
  \centering
  \includegraphics[width=0.5\textwidth]{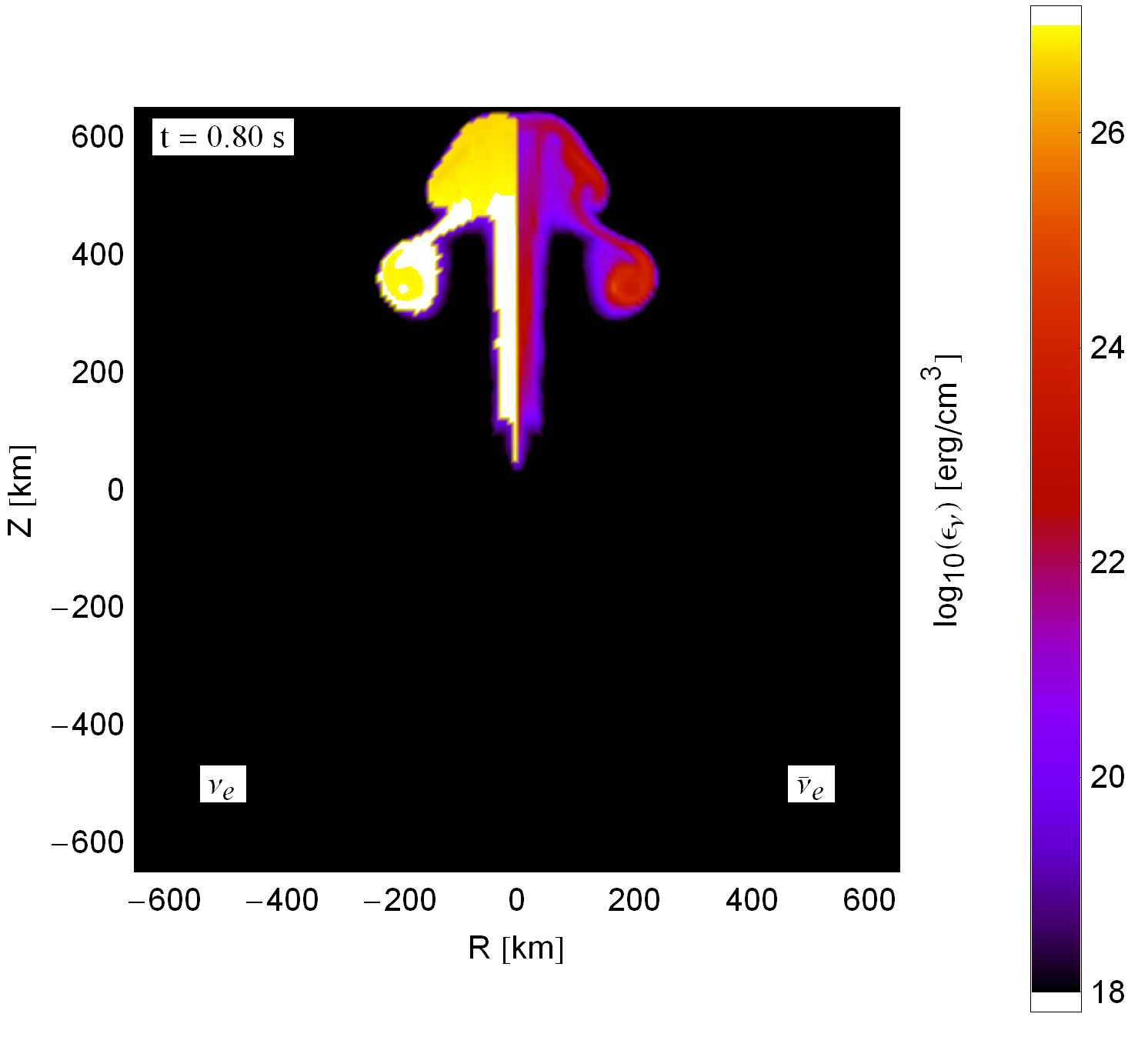}%
  \caption{Maps of the neutrino emissivity in the delayed-detonation
 model at t=0.8~s, i.e.\ near the peak of the neutrino emission
 produced by the initial failed deflagration stage;  left segment,
 $R < 0$ km $\nu_e$; right segment, $R > 0$ km $\bar{\nu}_e$.}
  \label{Lnu_y12_map_deflagration}
\end{figure}

\begin{figure}[bh!]
  \centering
  \includegraphics[width=0.5\textwidth]{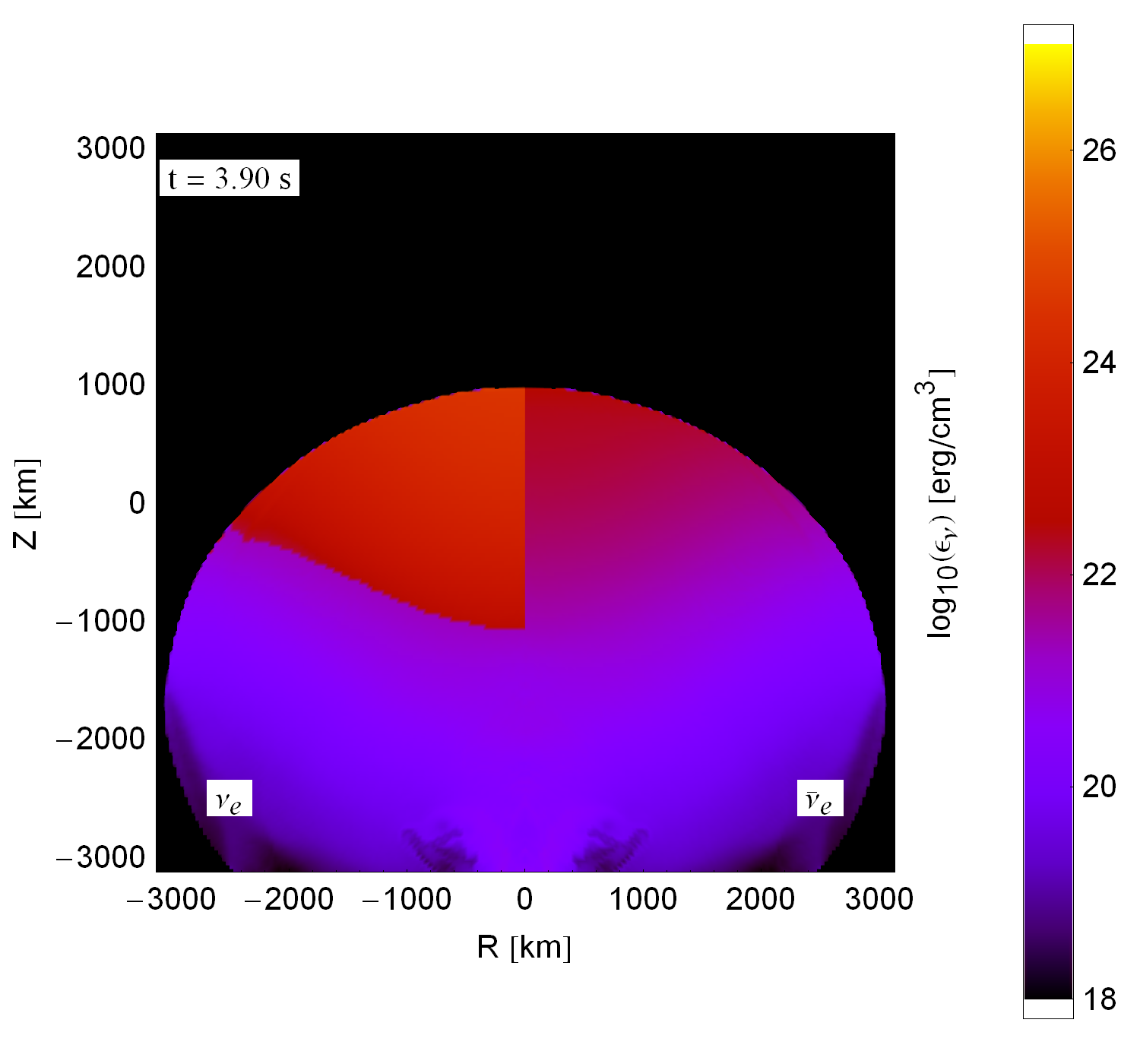}%
  \caption{Maps of the neutrino emissivity in the delayed-detonation
 model at t=$3.9$~s, i.e.\ near the peak of the neutrino emission
 produced by the detonation stage; left segment, $R < 0$ km
 $\nu_e$; right segment, $R > 0$ km $\bar{\nu}_e$.}
  \label{Lnu_y12_map_detonation}
\end{figure}

The electron flavor neutrino and antineutrino emission maps
(Figs.~\ref{Lnu_y12_map_deflagration} and \ref{Lnu_y12_map_detonation})
reflect the explosion physics. Roughly speaking, neutrino emission is
a by-product of the thermonuclear flame or the detonation wave. During
the deflagration stage, almost all $\nu_e$ are emitted in the
electron capture processes in the region incinerated by the thermonuclear
flame. Hot plumes expanding into the higher density gas are prominent
sources of electron neutrinos, because the electron capture rates are
increasing rapidly with the temperature (because of the thermal population of
the excited states with large matrix elements) and density (because of the 
Fermi-energy crossing capture threshold for excited nuclei). The total mass
involved in neutrino emission is much smaller than for pure
deflagration model, 0.2\% of the total white dwarf mass.

Antineutrinos ($\bar{\nu}_e$) are emitted from the much larger volume
heated by the thermonuclear burning. The electron antineutrino emission
from the thermal processes (pair annihilation) during the deflagration
stage is initially suppressed owing to the high degeneracy of the electron
gas. The main source of $\bar{\nu}_e$'s is pair annihilation,
Eq.~\eqref{pair}, and the reaction
$$ e^{+} + n \to p + \bar{\nu}_e.$$
After $t\approx 1$~s, pair annihilation
completely dominates the $\bar{\nu}_e$ flux (Fig.~\ref{Lnu_y12}b, red
solid curve).

The deflagration stage ends with a bubble breakout and the neutrino
emission from nuclear processes ends. Thermal neutrinos are still
emitted from the area heated during nuclear burning, but the neutrino flux
decreases by several orders of magnitude (see
Figs.~\ref{Lnu_y12}a-c). At $t=3.7$ s, the material accelerated by the
expanding bubble starts converging at the location opposite to the
bubble breakout point, and eventually triggers a
detonation. Interestingly, the thermal neutrino emission starts to
rise just before to the detonation ignition (Fig.~\ref{Lnu_y12}c). This
is because of the neutrino cooling of the colliding matter, which heats up
enough to produce $e^{+}e^{-}$ pairs. Once the detonation\footnote{The
detonation is a reactive wave in which a thin hydrodynamic shock
activates a thermonuclear burn and is followed by an extended post-shock
region in which the thermonuclear fuel is processed and the energy is
released \citep{fickett+79}.} is formed, the wave quickly moves into
the white dwarf core. The nuclear burning involves electron captures,
and weak nuclear neutrinos are the dominant component of the neutrino
emission (left segment in Fig.~\ref{Lnu_y12_map_detonation}).

In contrast to the pure deflagration, during the detonation phase a
large fraction of the white dwarf ($\approx 30$\% in mass) is
participating in producing the neutrino emission. We found that in
this case $\approx 50$\% of the emission is produced by matter with
$T_{NSE}<T_9<7.2$ and $7.85<\log_{10} \rho < 8.25$.  Thermal neutrinos
are also emitted from a much larger volume (of the deflagration-expanded
white dwarf) swept by the detonation (see right panel in
Fig.~\ref{Lnu_y12_map_detonation}), and they are the main contributor to
the $\bar{\nu}_e$ flux. Only residual pair neutrino emission from the
deflagration stage can still be seen at this time. Once the detonation
ends, however, the ejecta quickly expand and cool down adiabatically,
and the supernova becomes an exponentially fading source of thermal
neutrinos\footnote{See
\href{http://ribes.if.uj.edu.pl/snIa/}{http://ribes.if.uj.edu.pl/snIa/}
for step-by-step neutrino emissivity maps, animations, digitized
neutrino spectra, and additional data.}.

%
%
%

\subsubsection{Comparison of neutrino emission signatures}

One of the most exciting possibilities opened by the neutrino channel
is a potential for distinguishing between various explosion
scenarios.  While the overall number of scenarios is quite large, most
of them fit into either the pure deflagration or the delayed-detonation
category. Therefore, the two models analyzed in previous sections provide
a small but representative sample. We have at least three observables
available for the explosion diagnostics: the total energy radiated by
neutrinos (directly related to the observed number and energy of
events), the time variation of the neutrino signal (sensitive to the
burning speed and burning type), and the energy of detected neutrinos
(probing the degeneracy of the burning matter). The analyzed models differ
quite significantly in these three respects  (see Table~\ref{t:total}).
\begin{table*}[ht!]
\begin{center}
  \caption{Integrated properties of the model neutrino signals.
    \label{t:total}}
  \begin{tabular}{|c|c|c|c|c|}
    \hline
Model & n7d1r10t15c & Y12 (def) & Y12 (det) & Y12 (total) \\
    \hline
$E_{\nu}^{total}$ [erg] & $3.85 \times 10^{49}$ & $7.3 \times 10^{48}$ & $8.7 \times 10^{47}$ & $8.2 \times 10^{48}$ \\
$E_{\nu}^{total}/E_{nucl}^{total}$ & 0.03 & 0.05 & 0.0005  & 0.004 \\ 
    \hline
$E_{\nu_e}^{total}$ [erg]       & $3.85 \times 10^{49}$ & $7.3 \times 10^{48}$ & $7.7 \times 10^{47}$ & $8.05 \times 10^{48}$\\
$E_{\bar{\nu}_e}^{total}$ [erg] & $7.0 \times 10^{46}$   & $8.9 \times 10^{45}$ & $5.9 \times 10^{46}$ & $6.8  \times 10^{46}$\\
$E_{\nu_x}^{total}$ [erg]       & $6.4 \times 10^{46}$   & $2.2 \times 10^{45}$ & $4.4 \times 10^{46}$ & $4.6  \times 10^{46}$\\
    \hline
$\langle \mathcal{E}_{\nu_e} \rangle^{total}$ [MeV]       & 3.8 & 3.7 & 2.35 & 3.5   \\
$\langle \mathcal{E}_{\bar{\nu}_e} \rangle^{total}$ [MeV] & 2.9 & 3.0 & 1.9  & 2.0   \\
$\langle \mathcal{E}_{\nu_x} \rangle^{total}$ [MeV]       & 2.5 & 2.8 & 2.0  & 2.0   \\
    \hline
double $L_\nu$ peaks & no & peak 1 & peak 2 & yes \\
signal duration [s] & 1.0  & 1.0  & 0.4 & separation $\sim$3 sec \\

\hline
  \end{tabular}
\end{center}
\end{table*}
The most striking difference is the total emitted neutrino energy,
which almost entirely comes from the electron flavor neutrino. The
delayed-detonation model produces five times less energy in neutrinos
despite a comparable explosion energy. Therefore, if we look at a nearby
explosion that is unobscured by interstellar matter, we can easily identify
the explosion scenario provided the total (kinetic+radiative) explosion
energy can be determined. Neutrino energies are also a little bit
smaller in the delayed-detonation model (Table~\ref{t:total}).
Unfortunately, only $\nu_e$ provides a  clear signature. Other neutrino
flavors, including relatively easy to detect $\bar{\nu}_e$, are
emitted in comparable quantities. The total energy radiated in
$\nuebar$ is $\approx 7.0 \times 10^{46}$ erg for n7d1r10t15c,
comparable to $\approx 6.2 \times 10^{46}$ ergs for Y12. The average
$\nuebar$ energy in the Y12 model (3.5 MeV) is only 0.3 MeV
lower than a pure deflagration (3.8 MeV).

The characteristic double-peaked neutrino luminosity curve
(Fig.~\ref{Lnu_y12}) is a ``smoking gun'' of the delayed-detonation
supernova, although the second maximum is fairly weak. However, owing to
the $\approx 4$ seconds delay between the maxima, and compared to $\approx
2.5$ seconds long deflagration, a detection of a neutrino events a few
seconds apart would offer evidence for an explosion caused by a delayed
detonation.

\section{Discussion}

\subsection{Prospects for neutrino detection from a galactic type Ia supernova}

In the context of SN Ia neutrino emission, possibly the most important
question is whether the supernova neutrino signal can be measured
using the available neutrino-detection technologies. To answer this
question one requires the following information: (1) estimated
galactic supernova rates and expected supernova distances, (2) the
integrated supernova neutrino ($\nu_e$) and antineutrino
($\bar{\nu}_e$) spectra; (3) characteristics of suitable neutrino
detector. In the following discussion we will consider a supernova
located at the distance of 1~kpc\footnote{Before SN 1987A, it was not
unusual to adopt a 1~kpc distance to the ``future core-collapse
supernova;'' see, e.g., \cite{1984ApJ...283..848B}.}. The results for
a widely adopted 10~kpc distance (roughly a distance to the Galactic
Center with the corresponding volume including $\approx$50\% stars in
the Milky Way, \citealt{1980ApJS...44...73B}) can be obtained by
dividing the current numbers by a factor of $100$.

The selection of the interesting nuclei and processes of interest is
potentially quite complicated because of large number of the nuclei
involved in NSE neutrino emission, each with unique (often poorly known) 
spectral properties, and contribution from additional
thermal processes. To aid the selection process, we constructed a
diagram showing the temporal evolution of neutrino emission from
individual nuclides and/or processes integrated over the stellar
volume as a function of the neutrino energy.\footnote{Similar diagrams
can be used to discuss other phenomena, e.g., the evolution of
pre-supernovae \citep{Odrzywolek_SN1987A-20th}.} Specifically, we plot
$(\langle \mathcal{E}_\nu \rangle(t), F_\nu(t))$ on the
$F_\nu$-$\langle \mathcal{E}_\nu \rangle$ plane. This diagram might be
referred to as the $\nu$-HR diagram, with the mean neutrino energy
considered an analogue of the effective stellar temperature and the
neutrino flux now playing a role of the stellar bolometric
luminosity. For a given supernova distance and detector, one can also
show isocontours of detection rates. Because the knowledge of the mean
neutrino energy and integrated flux is not sufficient to reproduce the
energy spectrum, in calculating detection rates we are forced to
assume a single parameter spectral function. In neutrino astrophysics,
it is common to use the Fermi-Dirac function \citep{PhysRevD.41.2967}:
\begin{equation} \label{FD-spectrum} \Phi(\mathcal{E}_{\nu},t) \equiv
\frac{R(t)}{\langle \mathcal{E}_{\nu} \rangle(t)^3 } \frac{a \,
\mathcal{E}_{\nu}^2}{1+e^{b \, \mathcal{E}_{\nu}/\langle
\mathcal{E}_{\nu} \rangle}} \quad a\approx17.3574, \; b \approx
3.15137,
\end{equation}
where $R(t)$ is the integrated particle emission rate and $\langle
\mathcal{E}_{\nu} \rangle(t)$ the average neutrino energy dependent only
on time, and $a$ and $b$ normalize the spectrum.

\begin{figure*}
  \centering
  \includegraphics[width=\textwidth]{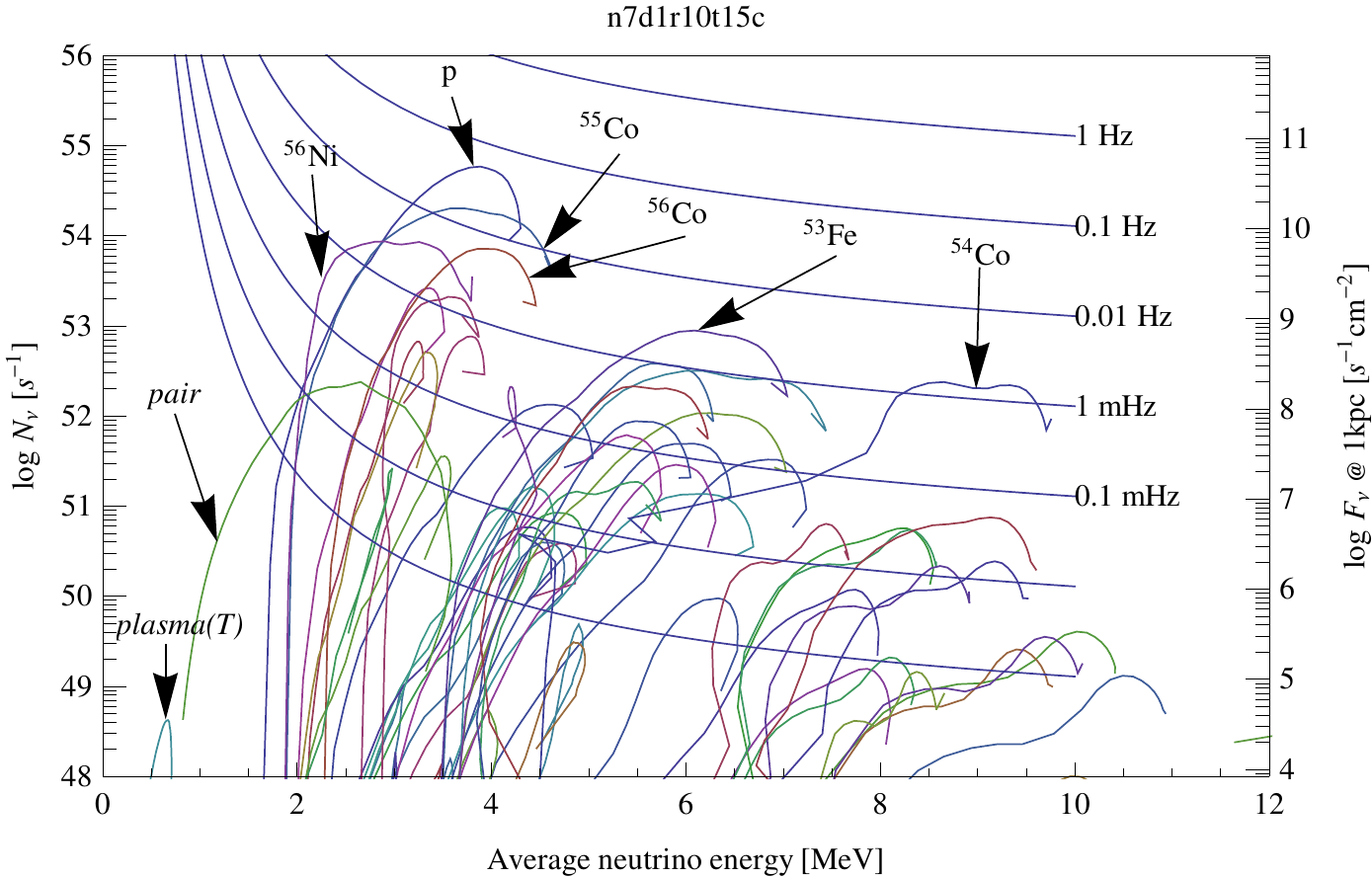}
  \caption{Neutrino-HR diagram for the n7d1r10t15c model. Every curve is a
    track on a $F_{\nu_e} - \langle\mathcal{E}_{\nu_e}\rangle$ plane
    produced by a single nucleus/ thermal process. Assuming a single
    parameter neutrino energy spectrum (Eq.~\eqref{FD-spectrum}), we
    are able to immediately select the most interesting for further
    analysis processes and estimate the expected signal in a given
    neutrino detection channel. Particularly, we  present
    detection of $\nu_e$ using elastic scattering off electrons with
    the threshold for detection of the electron kinetic energy of 4~MeV in
    a water Cherenkov detector.
    \label{nuHR-n7-ES4}}
\end{figure*}

For the assumed supernova distance of 1 kpc, the results for a pure
deflagration model and a Super-Kamiokande class detector (H$_2$O
target with the Cherenkov light detector with a threshold of 4~MeV) are
shown in Fig.~\ref{nuHR-n7-ES4}.
In particular, we conclude from the results
shown in Fig.~\ref{nuHR-n7-ES4}:
\begin{enumerate}
\item[(1)] the most important neutrino-producing nuclei for
Super-Kamiokande-like detector events terms are free protons and
\nuc{55}{Co}; the expected event rate is in 1~kt of H$_2$O up to 0.1/sec: because
an explosion takes $\approx$1 second in the Super-Kamiokande we expect up to $0.1/s/kt \times
32 kt \times 2\;\text{nuclei} \approx 6$ events from 1~kpc;
\item[(2)] secondary sources of detectable signal are: \nuc{56}{Ni},
\nuc{56}{Co}, \nuc{53}{Fe} and \nuc{54}{Co} with mean energies of
$\approx3$ MeV, $\approx 4$ MeV, $\approx 6$ MeV, and $\approx 9$ MeV,
respectively;
\item[(3)] numerous other nuclei as well as thermal processes produce
either a weak or an undetectable signal.
\end{enumerate}
Note that in Fig.~\ref{nuHR-n7-ES4} the evolution proceeds along
curves from high-energy to low-energy neutrinos (i.e. from right to
left). This is in contrast to core-collapse supernova neutrinos.

\begin{figure*}
  \includegraphics[width=\textwidth]{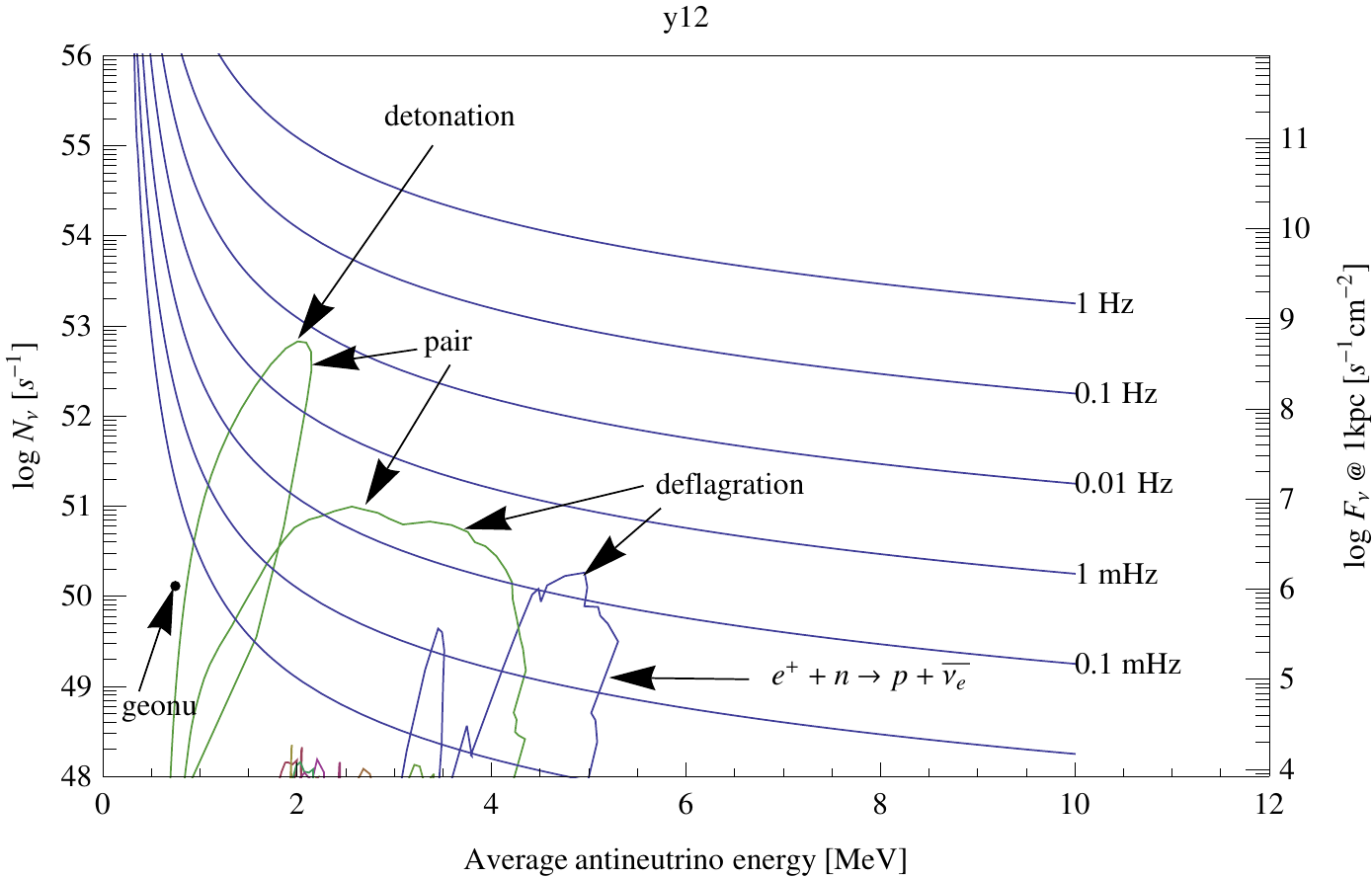}
  \caption{Antineutrino-HR diagram for Y12 model. Similar to
    Fig.~\ref{nuHR-n7-ES4}, but now we consider a detection of
    $\bar{\nu}_e$ via inverse beta decay in GdCl$_3$-loaded H$_2$O
    with a threshold of 2~MeV.
    \label{nuebarHR-y12-IBD2}}
\end{figure*}

The results of similar analyses for antineutrinos from the
delayed-detonation model, Y12, are shown in
Fig.~\ref{nuebarHR-y12-IBD2}.
We consider the inverse beta decay ($\nuebar + p \to n + e^{+}$) as
the detection channel, and a Gd-loaded water Cherenkov detector
proposed by \cite{Gadzooks} or a liquid scintillator
detector, e.g.\ KamLAND \citep{2003PhRvL..90b1802E}. We note that here the detection method is
simply the inverse of the essential production process ($e^{+} + n \to p +
\nuebar$). The analysis of Fig.~\ref{nuebarHR-y12-IBD2} leads to the
following conclusions:
\begin{enumerate}
\item[(1)] most important for $\nuebar$ emission processes are pair-annihilation
and positron capture on neutrons
\item[(2)] weak nuclear processes from nuclei are negligible
\item[(3)] the expected event rate is very low ($\sim$few~mHz/kt@1~kpc ); 
at least a half-megaton detector is required to observe a single event
from 1~kpc.
\end{enumerate}

Following the analysis of the $\nu_e$, $\nuebar$ detection in other cases, we
selected five most promising SN Ia neutrino experiments:
\begin{enumerate} 
\item IBD2: inverse beta decay $\bar{\nu}_e + p \to n +
e^{+}$ utilized in a large 50 kiloton target liquid scintillator detector 
(e.g.\ LENA \citealt{2007JCAP...11..011A,Undagoitia2006283,Oberauer2005108}) or
Gd-loaded water detector (\citealt{Gadzooks}) with 1.8 MeV threshold
\item ES0: elastic scattering off electrons 
$\nu_e + e^{-} \to \nu_e + e^{-}$
in a large 50 kt liquid scintillator (LENA) assuming
$\approx$0.2 MeV threshold 
\item ES4: elastic scattering off electrons
$\nu_e + e^{-} \to \nu_e + e^{-}$ in the extremely large water Cherenkov detector
Memphys \citep{2007JCAP...11..011A,1742-6596-171-1-012020}, 
Titan-D \citep{2001hep.ex...10005S,2008JPhCS.136b2057S,2008arXiv0810.1959K}, 
LBNE W.C. \citep{1742-6596-203-1-012079} etc. assuming
a standard 4.0~MeV detection threshold for recoil electrons
\item LAr: neutrino absorption in 100 kt of liquid argon \cite[see
e.g.][GLACIER proposal]{1742-6596-171-1-012020} detected using
coincidence of electrons and delayed gammas ($\nue+\nuc{40}{Ar} \to
\nuc{40}{K}^\ast+e^-$, \citealt{PhysRevD.34.2088}) and elastic
scattering off electrons ($E_\mathrm{th}=5$~MeV)
\item PES: elastic scattering off
protons in an advanced extremely low-background liquid scintillator detector 
like Borexino \citep{Alimonti2009568}
\item COS: coherent elastic scattering off high A nuclei (e.g.\ $^{72}$Ge) 
in a detector with a  threshold on the order of 100 eV.
\end{enumerate}
While scenarios IBD2, ES0, ES4, and LAr use a proven technology
\citep{2010JPhCS.203a2077F}, proton elastic scattering (PES) and
neutrino-nucleus coherent scattering (COH) have never been used in
practice for low $\nu$ energy. However, from theoretical analysis and
preliminary experimental results we expect to observe significant
progress in the development of neutrino detectors. Besides possible gains
from the development of advanced detection methods, larger target masses
are required for successful detection of SN Ia neutrinos in the
foreseeable future.

Table~\ref{t:nu_events}
\begin{table*}[ht!]
\begin{center}
\caption{Expected number of events triggered in the select proposed
neutrino detectors by a thermonuclear supernova located at a distance
of 1 kpc.
\label{t:nu_events}}
\begin{tabular}{|l|c|ccc|c c|}
\hline

\multirow{3}{*}{detector} & n7d1r10t15c  & \multicolumn{3}{c|}{Y12} & \multirow{3}{*}{proposals} &  \multirow{3}{*}{status} \\
                          & deflagration & deflagration & detonation & total \\
                          & 0-2.5s       & 0-2s         & 3.5-4.5s   & 0-7s \\
\hline 
ES4 (0.5 Mt)              & 19           & 3.2          & 0.1        & 3.3  & Hyper-Kamiokande, Memphys & under construction \\
LAr (100 kt)              & 21.4 + 1.5   & 3.8+0.24     & 0.08+0.005 & 3.9+0.25 & Glacier & under construction \\
IBD2 (50 kt)              & 0.2          & 0.01         & 0.06       & 0.07 & Gadzooks!, LENA & proposed \\
ES0 (50 kt)               & 14           & 2.7          & 0.26       & 2.9  & LENA & proposed \\
PES (50 kt)               & 60           & 11.1         & 0.8        & 12.0 & LENA & proposed \\
COH (1000 kg)             & 0.03         & 0.005        & 0.0003     & 0.006 & - & planned \\
\hline 
\end{tabular}
\end{center}
\end{table*}
shows the expected number of neutrino events for prospective neutrino
experiments. For a delayed-detonation, we separated the
contributions from the initial deflagration and the following delayed
detonation. For weak neutrinos and antineutrinos, the total number of
expected events is simply the sum of events produced in individual
explosion stages. For thermal neutrinos, there is also a minor
contribution from the neutrinos emitted during the period that separates
the two explosion stages and during the final expansion
stage. Clearly, the largest yield comes from the $\nu_e$ emission from
electron captures during the deflagration stage. This is expected
because the neutrino luminosity is dominated by these neutrinos and
reaches $1.1 \times 10^{49}$ erg/s for delayed-detonation and $6.4
\times 10^{49}$ for pure deflagration. Finally, the time delay between
the two emission maxima of a delayed-detonation SN and their relative
length will be very important aspects of the data analysis.

\subsubsection{Neutrino background and signal-to-noise ratio}

%
%
%
\begin{figure*}[ht!]
  \centering
  \includegraphics[width=0.95\textwidth]{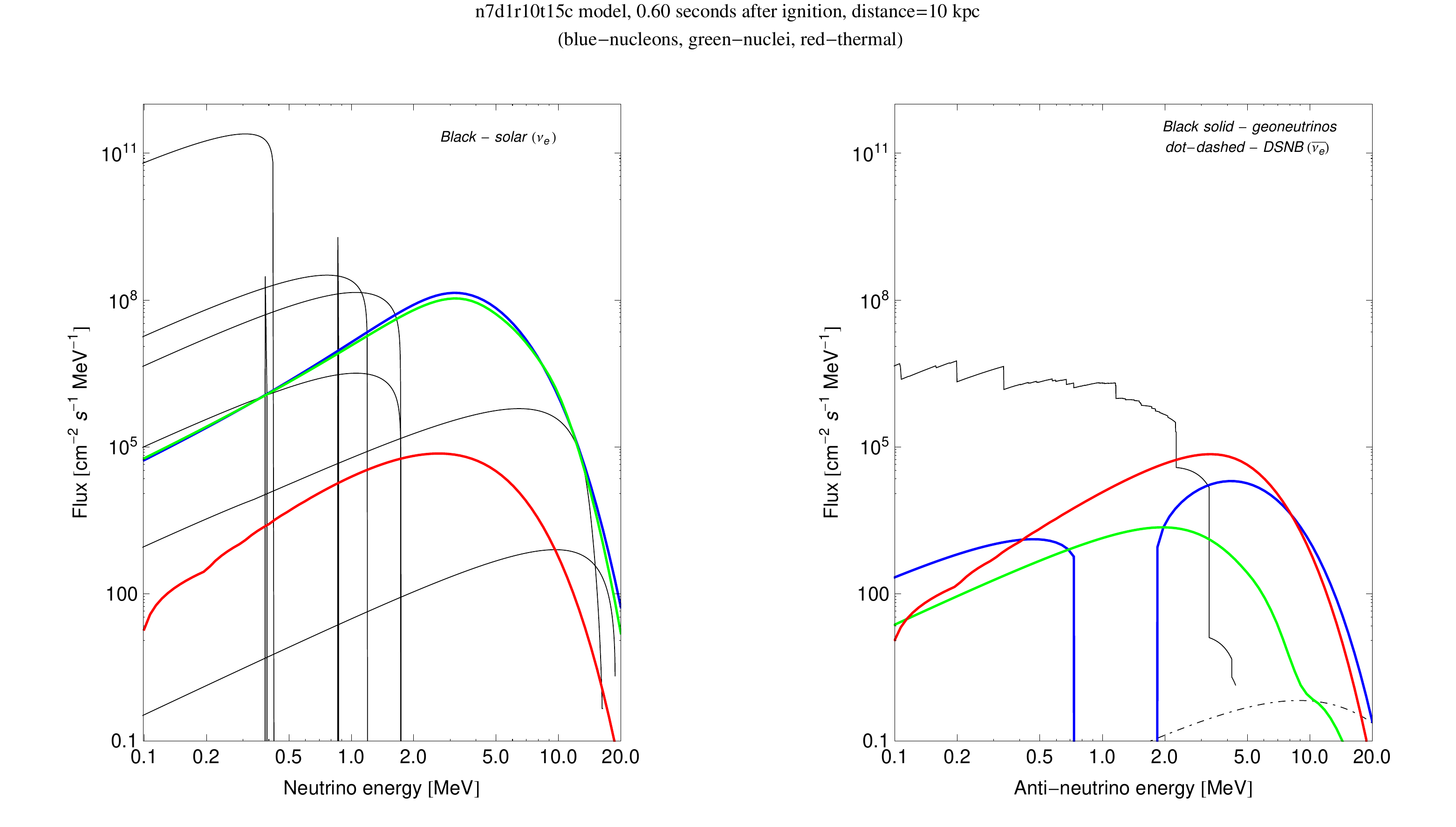}
  \caption{The $\nu_e$ (left) and $\bar{\nu}_e$ (right) model spectra
             of a pure deflagration supernova near the maximum of the
             neutrino emission and other recently studied sources. The
             supernova emission level is for an event located at a
             distance $d=10$ kpc. References for the data used: solar
             neutrinos, \cite{1538-4357-621-1-L85}; geoneutrinos at
             Kamioka, Japan, \cite{Enomoto_PhD,2006EM&P...99..131E};
             DSNB, \cite{lien+10}.
    \label{backgrounds}
    }
\end{figure*}
Additional comments on the expected background signals are due. For a 
$\nu_e$ emission and supernova at larger ($>$10~kpc) distance,
we face a problem of the background emission from \nuc{8}{B},
\nuc{7}{Be} and CNO solar neutrinos (left panel in
Fig.~\ref{backgrounds}). Here a directional detection could be a
solution, but no practical method of this kind exists. Electron
antineutrino emission will be blended with the geoneutrinos
(Fig.~\ref{backgrounds}, right) and the terrestrial nuclear power
plants. The geoneutrino flux varies slightly across the continental crust
and is much lower on the ocean floor
\citep{2006EM&P...99....1L,2005Natur.436..499A,2010PhLB..687..299B}.
Flux from human-made sources strongly depends on the location of the
detector and varies in time \citep{2005CRPhy...6..749L}. Other sources
of neutrinos, e.g.\ from cosmological core-collapse
supernovae\footnote{Those supernovae are a source of the diffuse
supernova neutrino background \citep{PhysRevD.79.083013}. The fact
that the sky is relatively dark in $\nuebar$, compared to individual
sources is the neutrino version of the Olbers paradox.} (flux $\ll
10~\cmms$, \citealt{lien+10}, \citealt{TotaniSato}; dot-dashed curve
in the right panel of Fig.~\ref{backgrounds}) are far below the
expected signal from a galactic SN Ia.  Relic neutrino flux is on the
order of $56\,c \simeq 1\times 10^{12}~\cmms$, but the energy is very
small in this case ($\sim10^{-4}$~eV).

From Fig.~\ref{backgrounds} it is clear that the neutrinos from a
galactic SN Ia could be detected, especially for the pure deflagration
event. Neutrino observations of such a supernova are mainly a
technological challenge (requires a very large detector mass, new
detection techniques, low-energy threshold, etc.) and, similar to SN
1987A, a matter of chance. \cite{1997A&A...322..431C} estimated
$4\pm1$ type Ia supernovae per millennium for Galaxy. An
Earth-centered ball with the radius of 10~kpc (1~kpc) contains
$\approx 50$\% ($\approx 0.5$\%) of stars \citep{1980ApJS...44...73B},
and the corresponding SN Ia explosion probability within a period of
10 years is therefore $\approx 0.02$ ($\approx 2 \times 10^{-4}$).

\section{Conclusions}

We have obtained and analyzed neutrino light curves and neutrino
spectra for two models of the most popular type Ia supernova explosion
scenarios: a pure deflagration and a delayed detonation. We 
discussed the role of physical conditions in producing neutrinos in
these types of explosions. In particular, the neutrino emission
studies allow us to directly probe the density, temperature, and
composition of the neutrino-emitting matter. This motivates
the development of neutrino experiments for exploring stellar evolution
physics beyond core-collapse supernova and solar applications.

Because of their cosmological importance and because their exact origins
remain unknown, thermonuclear supernovae are a class of exciting
future targets of the neutrino astronomy.  The upcoming challenge is a
detection of the SN Ia neutrinos. Several recently proposed neutrino
experiments will offer a sensitivity that will allow detecting a
thermonuclear event at kpc distances. More importantly, we find the
that \emph{the next generation of neutrino detectors will be able to
unambiguously identify the mechanism responsible for the
explosion}. In particular, SN Ia supernova electron neutrinos probe
the thermonuclear deflagration stage, while the electron antineutrinos
probe the detonation phase. Because the electron neutrinos stem almost
exclusively from electron captures associated with the thermonuclear
flame, they offer a means to study both nuclear and combustion physics
under extreme conditions. On the other hand, the delayed electron
antineutrino signal provides direct evidence for thermonuclear
detonation. Finally, the muon neutrinos are exclusively produced in
thermal processes and could potentially be used to extract weak
nuclear signals.

Given a relatively low neutrino luminosity of SN Ia events that are caused by
delayed detonations, their characteristic double-peaked neutrino light
curves can be used to reduce the false-alarm rate and serve as an
early warning system for this type of events. A pure deflagration SN
Ia produces only a single neutrino emission maximum with a somewhat
faster rise time compared to a delayed detonation. The predicted
number of observed neutrino events is, however, higher for deflagrations
thanks to both a higher neutrino luminosity and slightly higher energies
of the emitted neutrinos. For a $0.5$~Mt classical water Cherenkov
detector (LBNE WC, long baseline neutrino experiment
\citealt{1742-6596-203-1-012079}; Memphys
\citep{2007JCAP...11..011A,1742-6596-171-1-012020}), we predict
the recording about 20 elastic scattering events above 4~MeV per second
for a SN Ia event located at a distance of 1 kiloparsec. Still larger
detectors (e.g.\ Titan-D
\citealt{2001hep.ex...10005S,2008JPhCS.136b2057S,2008arXiv0810.1959K})
almost certainly guarantee positive detection of a galactic SN
Ia. However, this holds true only for a pure deflagration; \emph{the
predicted neutrino fluxes for a delayed detonation are about five times
lower which makes these events much harder to detect}. We also found that
the neutrino emission is very similar to two-dimensional axisymmetric
and spherically symmetric pure deflagration models (i.e. W7 by
\cite{Nomoto_W7}). This leads us to believe that \emph{the neutrino
observations will not help to distinguish between specific scenarios
of pure deflagrations} (e.g.\ ignition occurring at a single point or
at multiple points).

The majority of neutrino experiments considered here (detectors I, II,
and IV in Table \ref{t:nu_events}) use large amounts of a liquid
scintillator.  This type of experiments might be the most viable and
successful in detecting type Ia supernovae, especially if a proton
elastic scattering (PES) method is used \citep{Beacom_PES}. One
example of such a device is the Borexino detector
\citep{Alimonti2009568}. Although it is perhaps too small for detecting a
thermonuclear supernova at a kpc distance, Borexino will be an
essential testbed for the proposed and much larger LENA
\citep{2007JCAP...11..011A,Undagoitia2006283,Oberauer2005108} and
other similar experiments \citep{1742-6596-203-1-012137}. We also note
that neutrinos can be detected through neutrino-nucleus elastic
scattering
\citep{PhysRevD.30.2295,1748-0221-3-09-P09007,Collar_group,barbeau+03}. However,
a practical application of this technique to SN Ia may not be possible
because of the prohibitively large required mass of the detector.

We conclude that a significant progress in terms of neutrino detection
methods is needed for the neutrinos to become a practical tool for
studying type Ia supernovae. However, a detection of a thermonuclear
event at a distance of few kiloparsecs will be within the reach of the
planned neutrino observatories and will offer a perfect chance to
identify the mechanism that drives the explosion.

\begin{acknowledgements}
We thank Thomas Janka for encouragement and helpful advice, and an
anonymous referee for comments that helped improving the initial
version of this paper. TP was supported through the DOE grant
DE-FG52-03NA000064. This research used resources of the National
Energy Research Scientific Computing Center, which is supported by the
Office of Science of the U.S. Department of Energy under Contract No.\
DE-AC02-05CH11231, and NASA's Astrophysics Data System.
\end{acknowledgements}
%
%
%
%
%
\bibliographystyle{aa}
\bibliography{15133}
%
%
%


\Online


\begin{figure*}[ht!]
  \centering
  \includegraphics[width=0.95\textwidth]{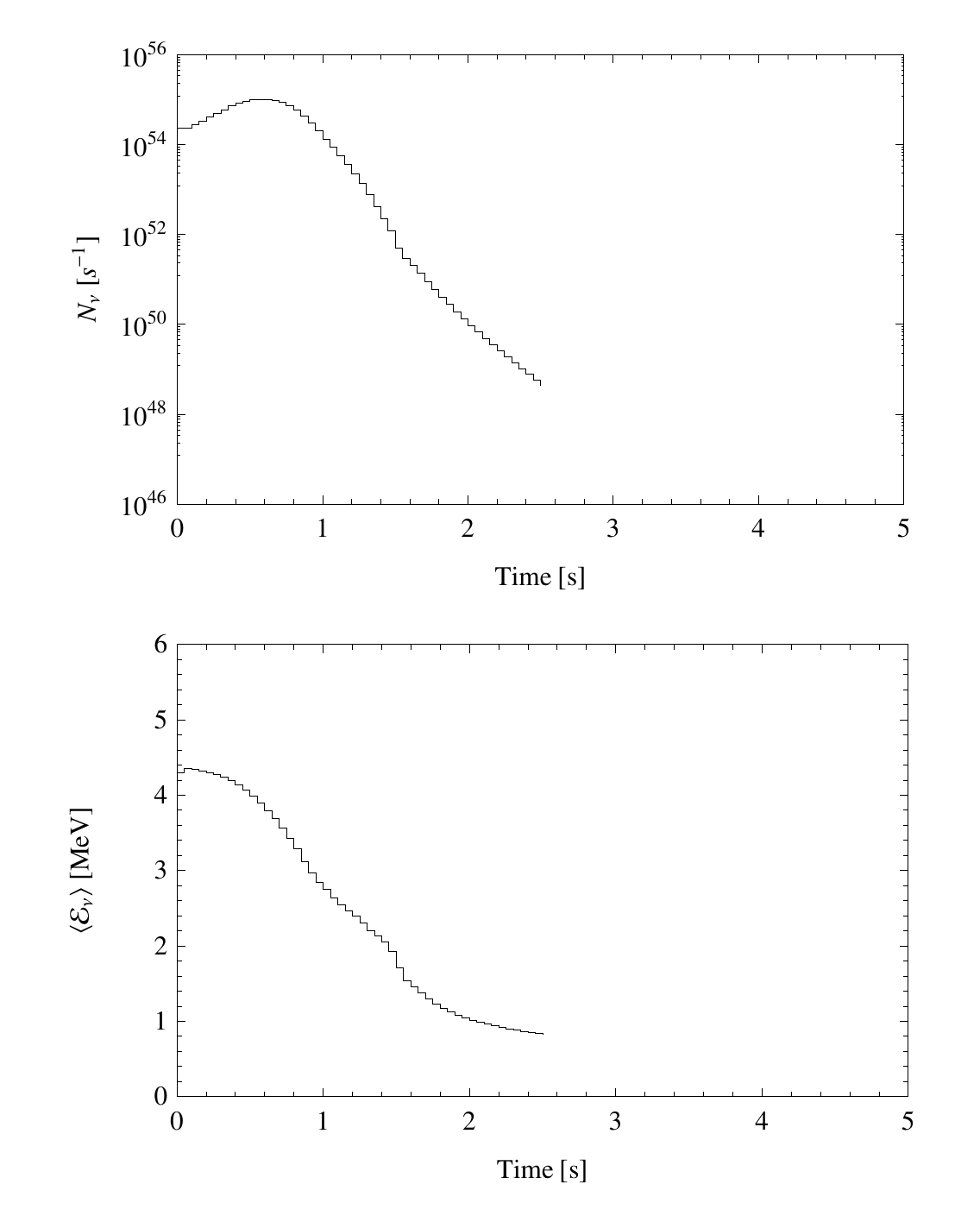}
  \caption{Model neutrino ($\nue$) particle emission 
    in the deflagration model n7d1r10t15c (top). Average neutrino energy (bottom).
    \label{NE1_n7}
    }
\end{figure*}

\begin{figure*}[ht!]
  \centering
  \includegraphics[width=0.95\textwidth]{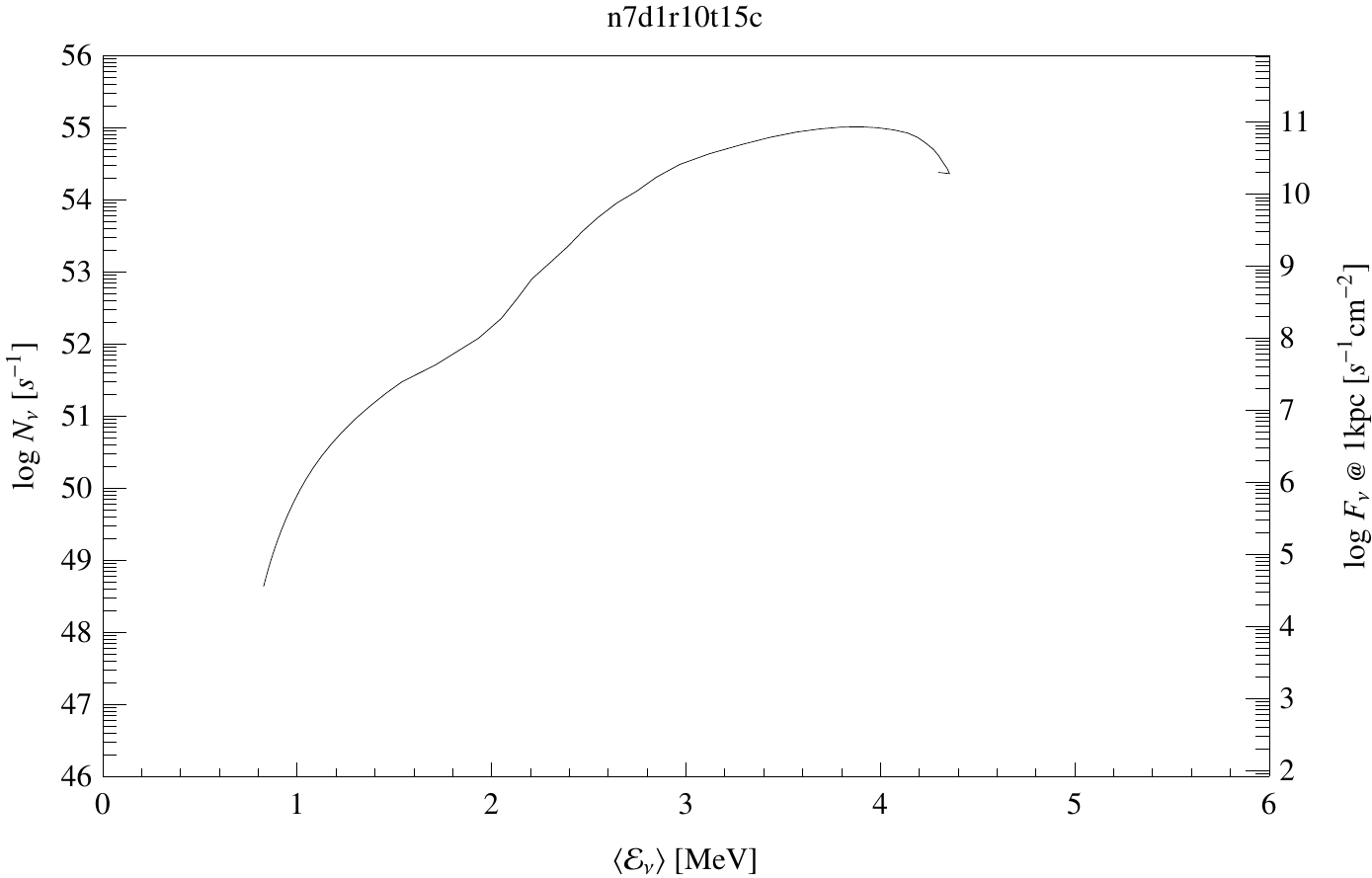}
  \caption{Total $\nue$-HR diagram for the deflagration model n7d1r10t15c.
    \label{NE1_n7a}
    }
\end{figure*}

\begin{figure*}[ht!]
  \centering
  \includegraphics[width=0.95\textwidth]{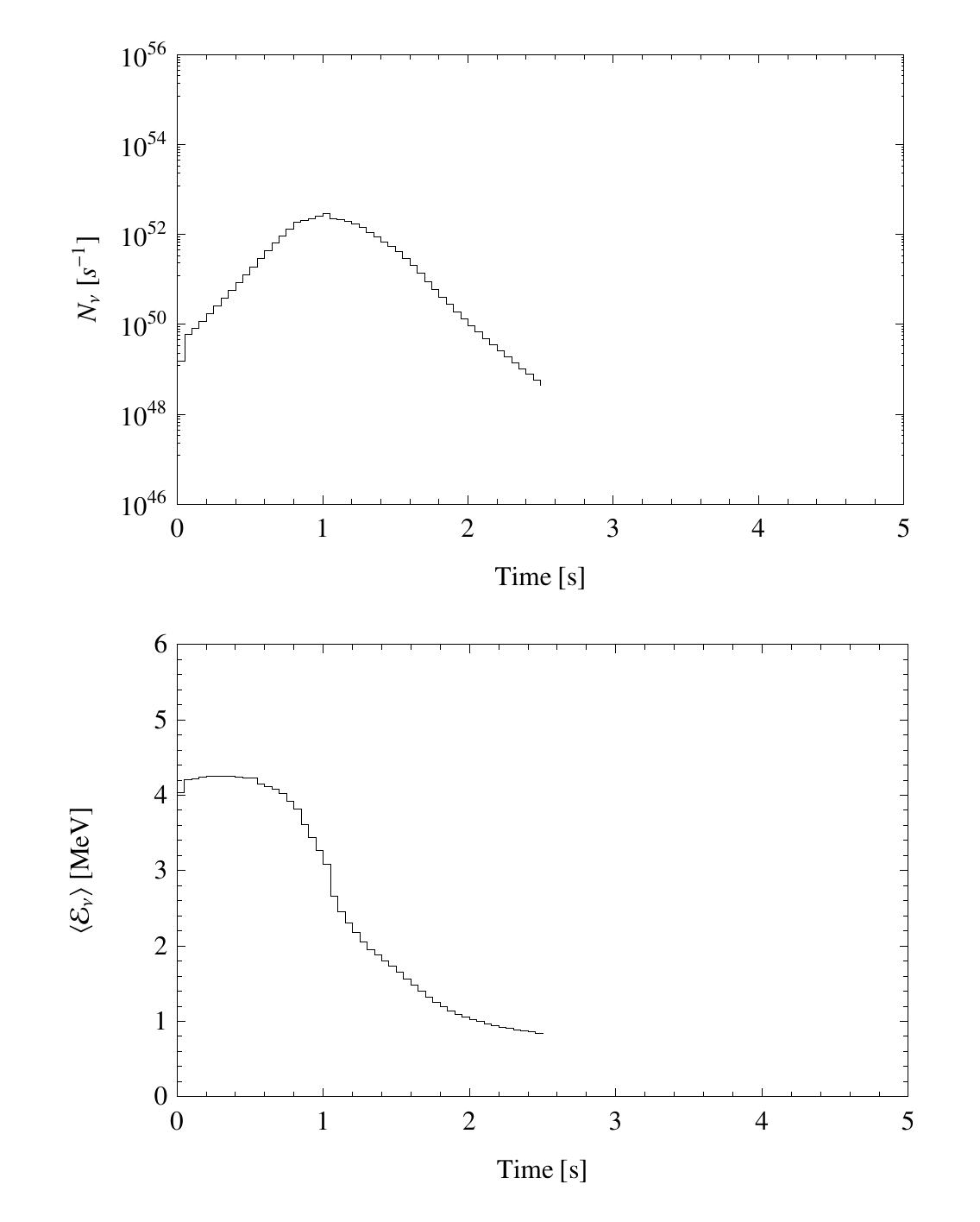}
  \caption{Model antineutrino ($\nuebar$) particle emission 
    in the deflagration model n7d1r10t15c (top). Average antineutrino energy (bottom).
    \label{NE2_n7}
    }
\end{figure*}

\begin{figure*}[ht!]
  \centering
  \includegraphics[width=0.95\textwidth]{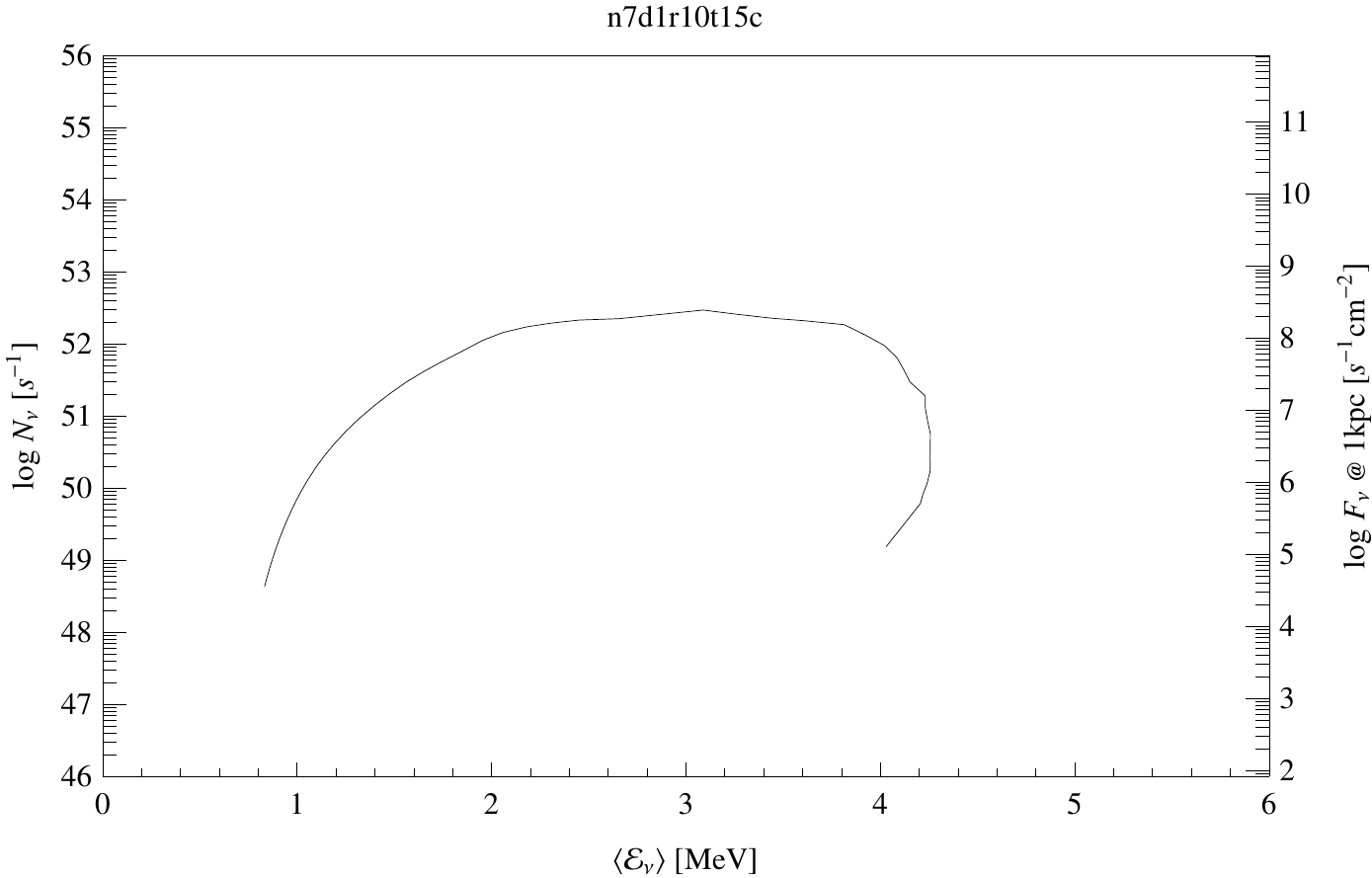}
  \caption{Total $\nuebar$-HR diagram for the deflagration model n7d1r10t15c.
    \label{NE2_n7a}
    }
\end{figure*}

\begin{figure*}[ht!]
  \centering
  \includegraphics[width=0.95\textwidth]{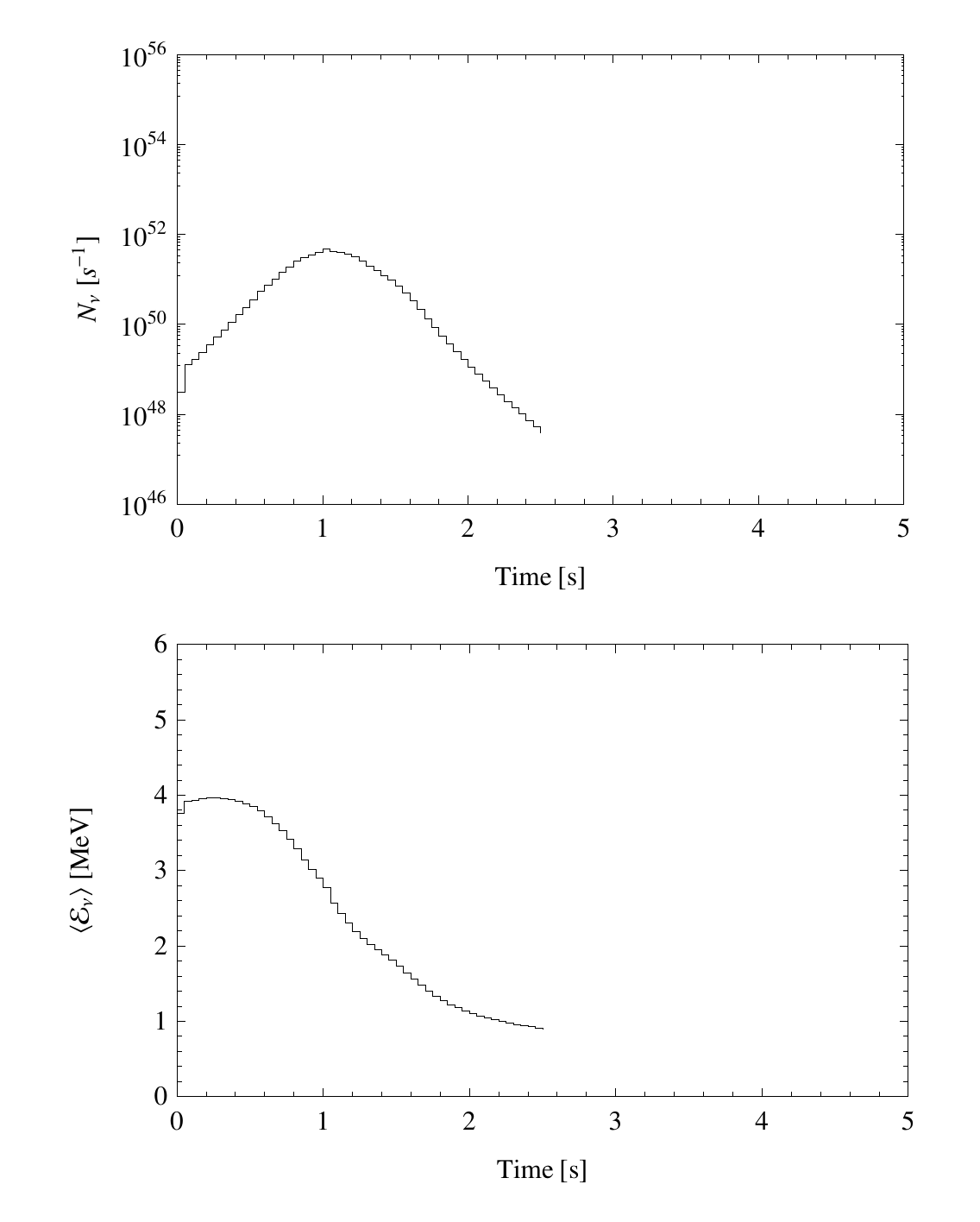}
  \caption{Model muon/tau neutrino ($\numu$) particle emission 
    in the deflagration model n7d1r10t15c (top). Average neutrino energy (bottom).
    \label{NE3_n7}
    }
\end{figure*}

\begin{figure*}[ht!]
  \centering
  \includegraphics[width=0.95\textwidth]{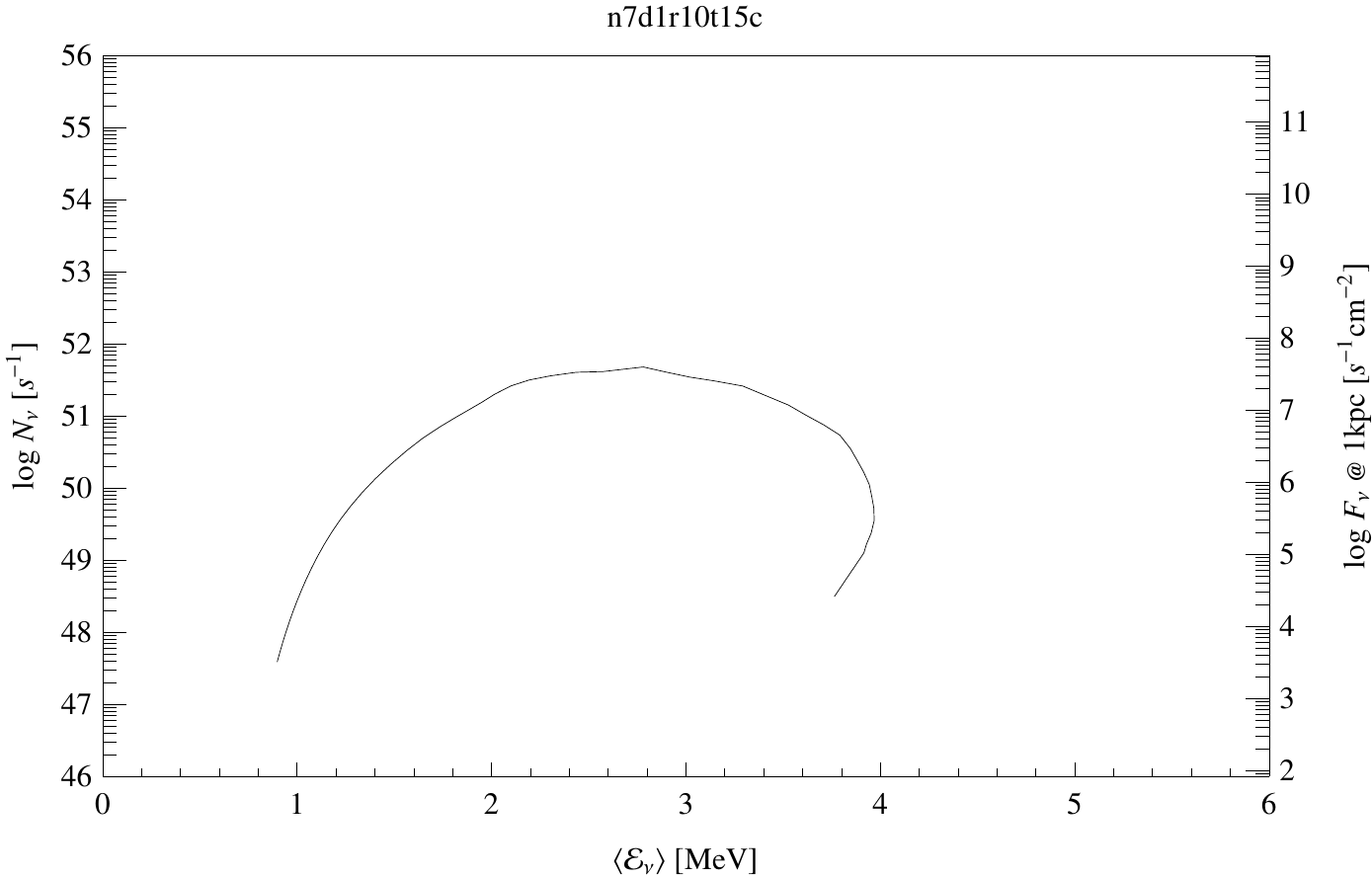}
  \caption{Total $\numu$-HR diagram for the deflagration model n7d1r10t15c.
    \label{NE3_n7a}
    }
\end{figure*}


\begin{figure*}[ht!]
  \centering
  \includegraphics[width=0.95\textwidth]{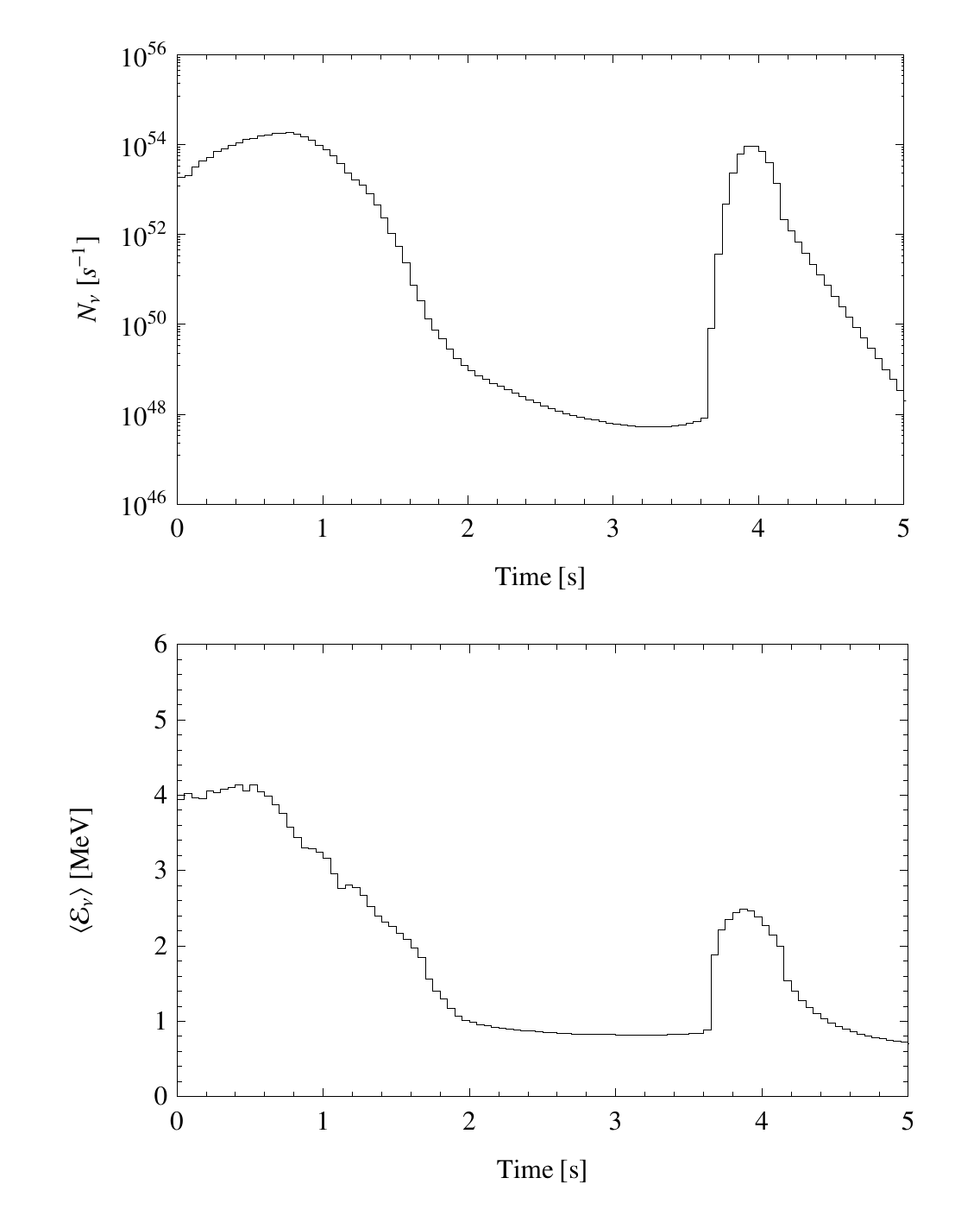}
  \caption{Model neutrino ($\nue$) particle emission 
    in the delayed detonation model Y12 (top). Average neutrino energy (bottom).
    \label{NE1_y12}
    }
\end{figure*}

\begin{figure*}[ht!]
  \centering
  \includegraphics[width=0.95\textwidth]{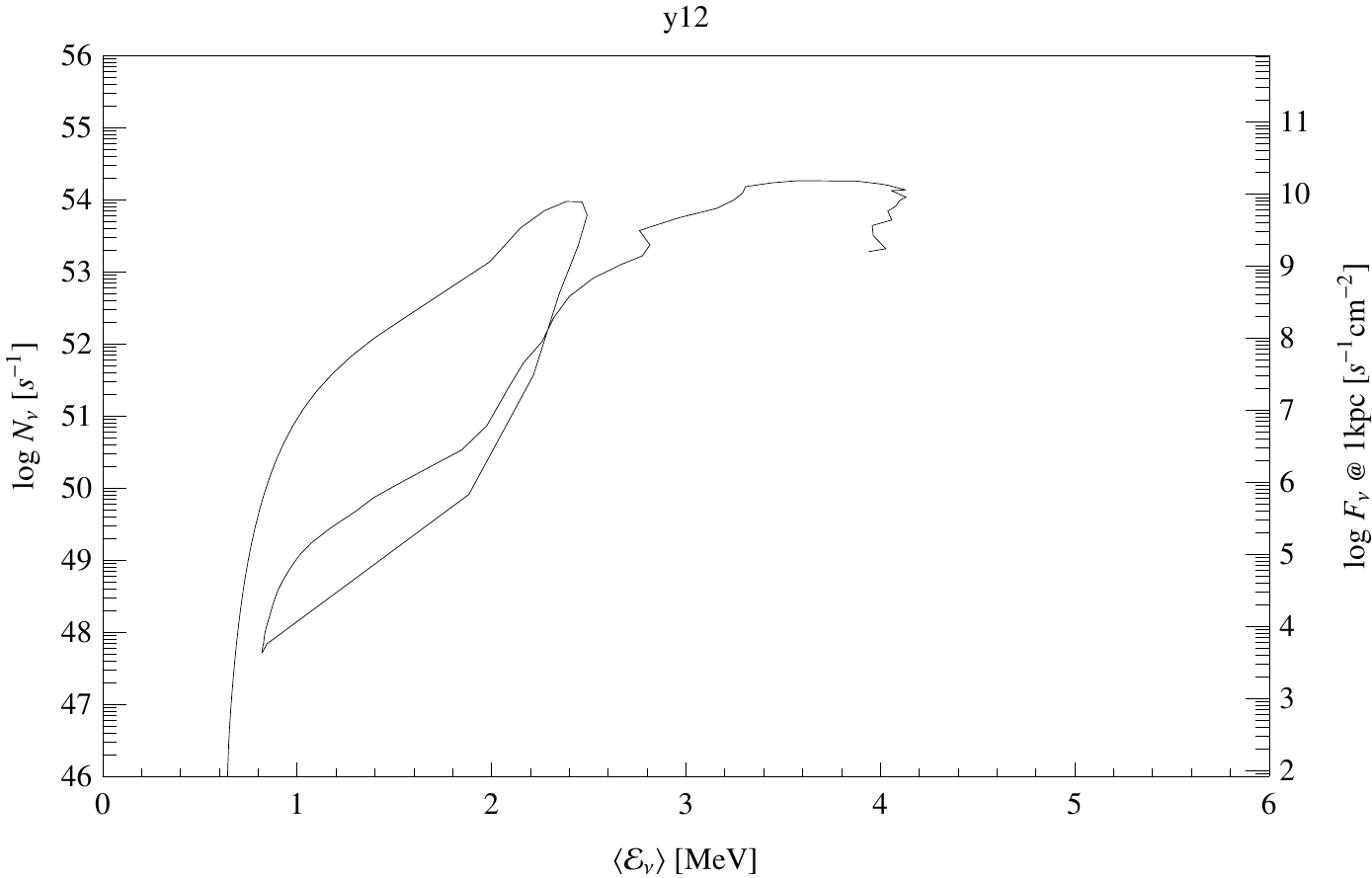}
  \caption{Total $\nue$-HR diagram for the delayed detonation model Y12.
    \label{NE1_y12a}
    }
\end{figure*}

\begin{figure*}[ht!]
  \centering
  \includegraphics[width=0.95\textwidth]{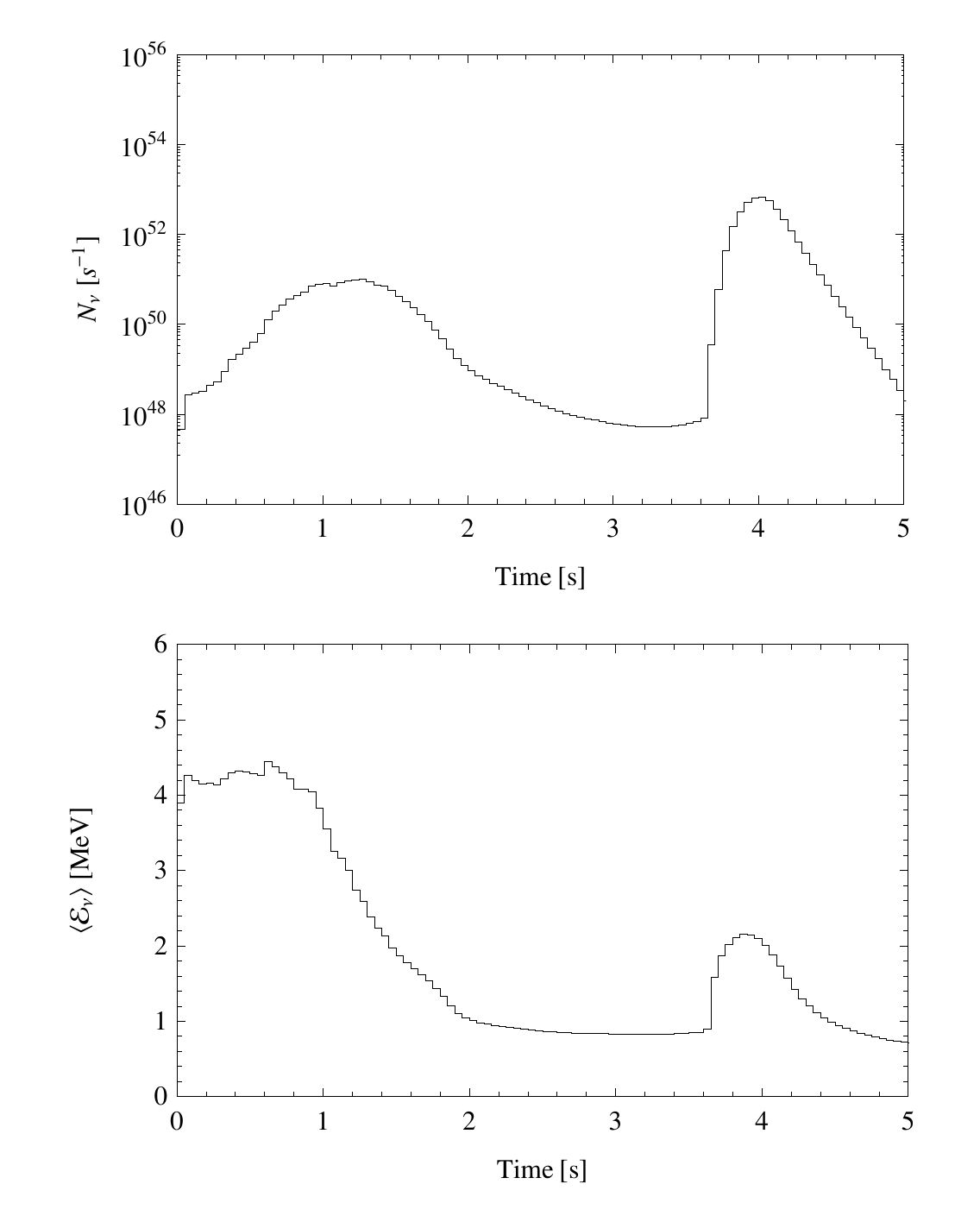}
  \caption{Model antineutrino ($\nuebar$) particle emission 
    in the delayed detonation model Y12 (top). Average antineutrino energy (bottom).
    \label{NE2_y12}
    }
\end{figure*}

\begin{figure*}[ht!]
  \centering
  \includegraphics[width=0.95\textwidth]{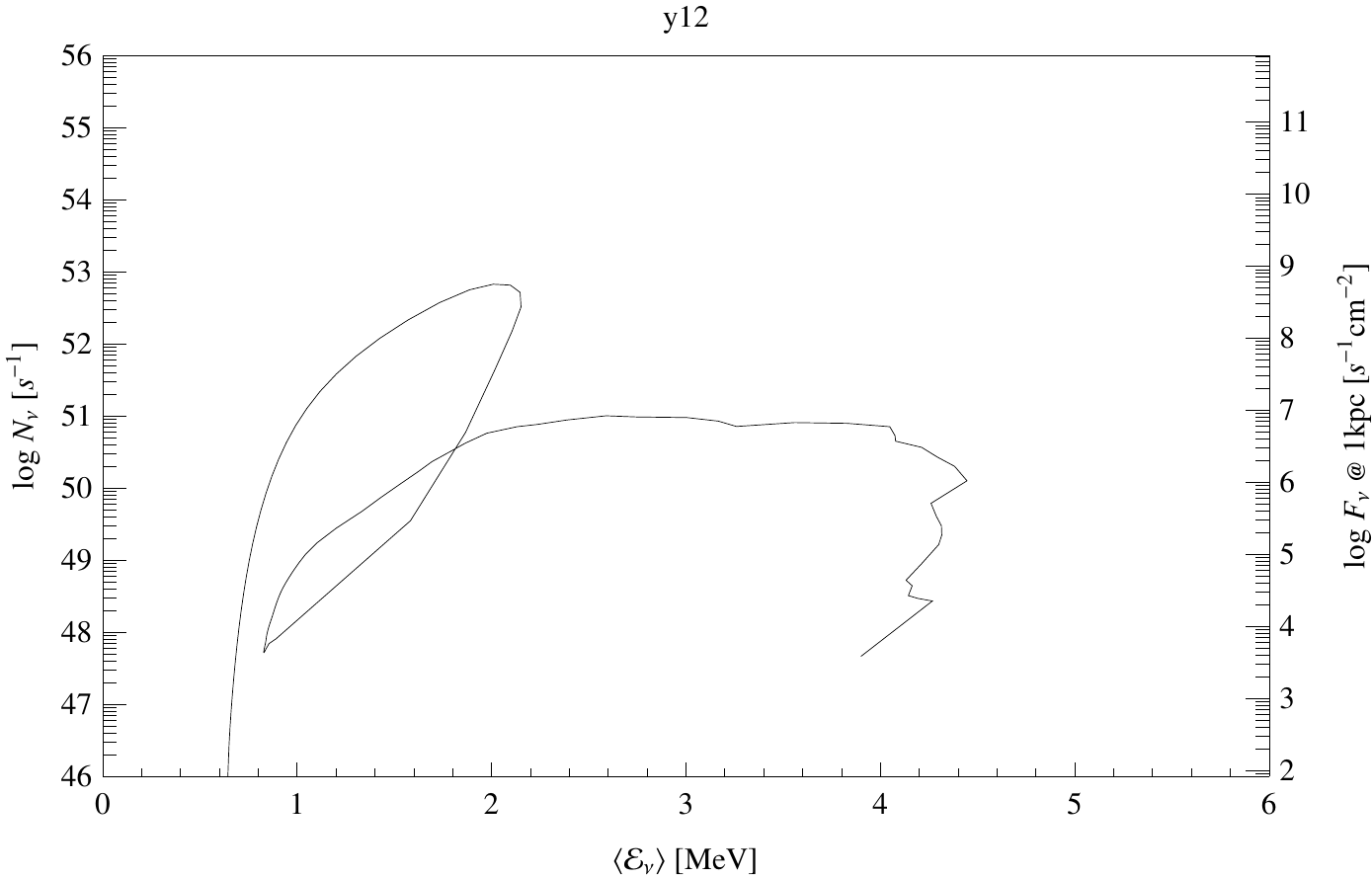}
  \caption{Total $\nuebar$-HR diagram for the delayed detonation model Y12.
    \label{NE2_y12a}
    }
\end{figure*}

\begin{figure*}[ht!]
  \centering
  \includegraphics[width=0.95\textwidth]{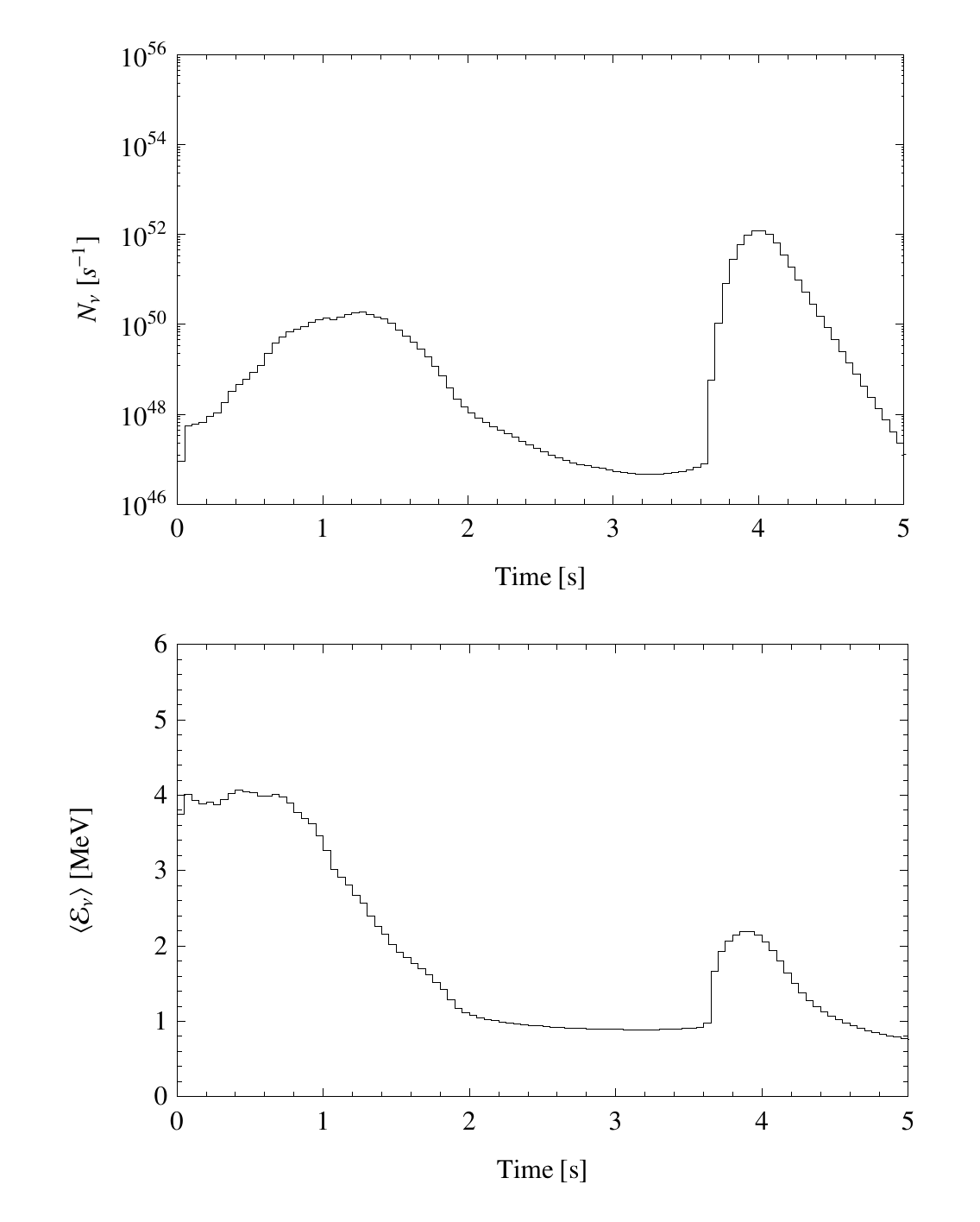}
  \caption{Model muon/tau neutrino ($\numu$) particle emission 
    in the delayed detonation model Y12 (top). Average neutrino energy (bottom).
    \label{NE3_y12}
    }
\end{figure*}

\begin{figure*}[ht!]
  \centering
  \includegraphics[width=0.95\textwidth]{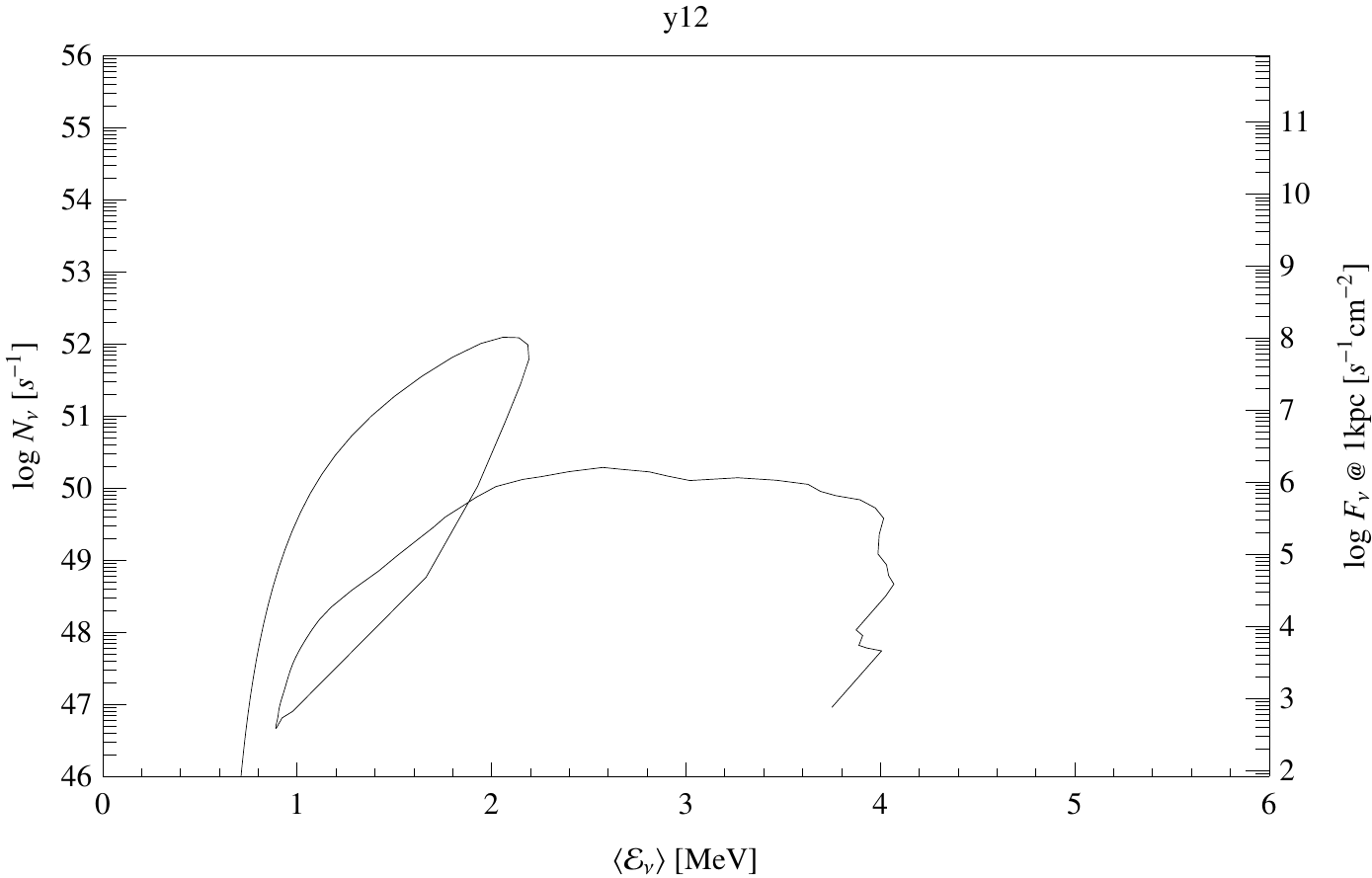}
  \caption{Total $\numu$-HR diagram for the delayed detonation model Y12.
    \label{NE3_y12a}
    }
\end{figure*}

\end{document}